\documentclass{aa}

\usepackage{scalefnt}
\usepackage[]{aalongtable}
\usepackage{graphicx}
\usepackage{epstopdf}
\usepackage[varg]{txfonts}
\begin{document}

   \title{First stars \\ XVI. STIS/HST abundances of
 heavy-elements in the uranium-rich star CS 31082-001\thanks{Based on observations made with the NASA/ESA Hubble Space 
Telescope (HST) through the Space Telescope Science Institute, operated by
the Association of Universities for Research in Astronomy, Inc., under NASA
contract NAS5-26555; and with the ESO Very Large Telescope at Paranal Observatory, 
Chile; Progr. ID 165.N-0276.
} }

\author{
C. Siqueira Mello Jr.\inst{1,2}
\and
M. Spite\inst{2}
\and
B. Barbuy\inst{1}
\and
F. Spite\inst{2}
\and
E. Caffau\inst{3,2}
\and
V. Hill\inst{4}
\and
S. Wanajo\inst{5}
\and
F. Primas\inst{6}
\and 
B. Plez\inst{7}
\and
R. Cayrel\inst{8}
\and
J. Andersen\inst{9,10}
\and
B. Nordstr\"om\inst{9}
\and
C. Sneden\inst{11}
\and
T.C. Beers\inst{12}
\and
P. Bonifacio\inst{2}
\and 
P. Fran\c cois\inst{8}
\and
P. Molaro\inst{13}
}
\offprints{C. Siqueira Mello Jr., email: cesar.mello@usp.br).}

\institute{
Universidade de S\~ao Paulo, IAG, Rua do Mat\~ao 1226,
Cidade Universit\'aria, S\~ao Paulo 05508-900, Brazil
\and
GEPI, Observatoire de Paris,  CNRS, UMR 8111, F-92195 Meudon Cedex, France
\and
Zentrum f\"ur Astronomie der Universit\"at Heidelberg, Landessternwarte, K\"onigstuhl 12, 69117 Heidelberg, Germany
\and
Universit\'e de Sophia-Antipolis,
 Observatoire de la C\^ote d'Azur, CNRS UMR 6202, BP4229, 06304 Nice Cedex 4, France
\and
National Astronomical Observatory of Japan, 2-21-1 Osawa, Mitaka, Tokyo 181-
8588, Japan
\and
 European Southern Observatory, Karl Schwarzschild Strasse 2, 85748
Garching bei M\"unchen, Germany
\and
LUPM, CNRS, UMR 5299,
 Universit\'e de Montpellier II, F-34095 Montpellier Cedex 05, France
\and
GEPI, Observatoire de Paris, CNRS, UMR 8111, 61 Av. de l'Observatoire, 75014 Paris, France
\and
The Niels Bohr Institute, Juliane Maries Vej 30, DK-2100 Copenhagen, Denmark
\and
Nordic Optical Telescope, Apartado 474, ES-38700 Santa Cruz de La Palma, Spain
 \and 
University of Texas at Austin, Department of Astronomy, Austin, TX 78712, USA  
\and
Michigan State University, Department of Physics \& Astronomy,
and JINA: Joinst Institute for Nuclear Physics, East Lansing,
MI 48824, USA
\and
INAF - Osservatorio Astronomico di Trieste, via Tiepolo 11, I-34143 Trieste, Italy
}

   \date{Received July 5, 2012; Accepted November 8, 2012.}

% \abstract{}{}{}{}{} 
% 5 {} token are mandatory
 
  \abstract
{Detailed abundances of the elements produced by r-process nucleosynthesis in various circumstances are our best observational 
clues to their origin, since the site(s) of r-element production is(are) still not known with certainty. 
A small fraction of extremely metal-poor (EMP) stars exhibit excesses of heavy neutron-capture elements produced in 
the r-process, and CS 31082-001 is among the 4 well-known r-process-enhanced EMP stars. 
Observations with HST/STIS provide abundances for elements observable only from the UV region. The third peak elements 
were analyzed in a previous paper, and here we present a complete analysis of heavy elements based on the  near-UV spectra from 
HST/STIS and a new VLT/UVES spectrum.
}
{Here we aim to supplement the optical data with abundances from near-UV spectroscopy of the first and second peak of the r-elements, 
which are crucial to giving insight into the nucleosynthesis of the elements beyond iron. The UVES spectrum provided additional 
measurements, thereby improving the previous results.}
{The spectra were analyzed with the OSMARCS LTE model atmosphere and with a consistent approach based on the 
spectrum synthesis code Turbospectrum to derive abundances of heavy elements in CS 31082-001, using updated oscillator strengths 
from the recent literature. We computed synthetic spectra for all lines of the elements of interest, checking for proper intensities 
and possible blends. We combined the abundances of heavy elements derived in previous works with the derivation of 
abundances from all reliable new list of lines, 
for the first and second peaks of r-elements.}
{We were able to derive new abundances for 23 n-elements, 6 of them - Ge, Mo, Lu, Ta, W, and Re - were not available in 
previous works, making this star the most complete r-II object studied, with a total of 37 detections of n-capture elements. 
We also present the first NLTE+3D lead abundance in this star. The results provide improved constraints on the nature of the r-process.}
   {}


   \maketitle
%
%________________________________________________________________

\section{Introduction}

The origin of the elements beyond the iron peak has been described as the result of two major mechanisms of neutron capture, 
the s-process and the r-process. The s-process occurs at a slower rate than the half-life of beta decay, 
while the r-process occurs at a rapid rate, which is shorter than the beta decay intervention timescale \citep{Burbidge1957}. 
The difference in the timescale is associated with different neutron fluxes, which are used by the seed nuclei to 
build heavier nuclei, and as a consequence, completely different sites are needed to allow for these mechanisms to occur. 
However, the site(s) of r-element production is(are) still not known with certainty \citep[e.g.][]{Wanajo2006,Kratz2007,Thielemann2010}. 

The popular models involve high-entropy neutrino-driven winds of neutron-rich matter, 
which build up heavy nuclei near the neutrino sphere of a core-collapse supernova \citep[][and references therein]{Woosley1994,Wanajo2007}. 
Studies of the Galactic chemical evolution confirm the likelyhood of core-collapse supernovae, in particular near their 
low-mass end (8-10 M$_{\odot}$), as the dominant source of r-process elements \citep[e.g.][]{Mathews1990,Ishimaru1999}. 

However, hydrodynamical simulations with accurate neutrino transport show that neutrino winds are proton-rich \citep{Fischer2010,Hudepohl2010} 
or only slightly neutron-rich \citep{Martinez2012,Roberts2012,RobertsReddy2012} and
 not very neutron-rich as found in some older simulations \citep{Woosley1994}. 
Since hydrodynamical simulations still encounter difficulties reproducing
 the astrophysical conditions of the process, the neutrino wind scenario for the origin of 
the heavy r-process elements is doubtful. 

Neutron-rich ejecta from neutron star-neutron star (NS-NS) or black hole-neutron star (BH-NS) 
binary mergers have been suggested as plausible alternative astrophysical sites of the main r-process 
\citep{Lattimer1977,Meyer1989,Freiburghaus1999,Surman2008,Goriely2011,Wanajo2012,Korobkin2012}. 

As recently discussed in \citet{Peterson2011}, studies of the origin of the lightest trans-Fe elements, from gallium through 
cadmium (Z = 31 to 48), are even more complex. These elements have been attributed in varying degrees to 
a weak s-process \citep{Clayton1968,Kappeler1989}, to a so-called light element primary process \citep[LEPP;][]{Travaglio2004} such as a weak r-process 
\citep{WanajoIshimaru2006,Farouqi2010,Wanajo2011}, and/or to the $\nu$p-process \citep{Frohlich2006,Pruet2006,Wanajo2006,Arcones2011} in core-collapse supernovae.

Detailed abundances of the elements produced by r-process nucleosynthesis in various circumstances are our best observational 
clues to the nature of this mechanism, and a quite good picture can be obtained by considering the products of heavy-element 
production in the first generation(s) of stars, as recorded in the low-mass stars that survive until today. Actually, 
neutron-capture elements are present in some of the oldest stars, which are Galactic halo stars extremely metal-poor (EMP),
  and studies  of their abundance are 
being employed to provide clues to and constraints on the nature of the synthesis and the identities of the stellar generations 
in the early Galaxy, and also to provide new insight into the astrophysical site(s) for the r-process. 

CS 31082-001 is in the group of the 12 EMP r-II\footnote{
Following the classification from \citet{Beers2005}.} 
giant stars known, and it is one of the most extreme r-element enhanced giants. Actually, together with CS 22892-052 and HE 1523-0901, 
this star has attracted the most attention, and its abundance patterns have 
been studied in considerable detail. Recently, \citet{Barbuy2011} did a complete analysis of the third-peak r-process elements and actinides 
abundances in CS 31082-001 using near UV HST/STIS spectra.

In the present paper, we analyze the first- and second-peak r-elements from near UV lines in the same STIS spectra and a new 
UVES/VLT spectrum centered at 340 nm. The first peak comprises the elements 38$\leq$Z$\leq$48, 
while the second peak comprises 56$\leq$Z$\leq$72, including the lanthanides (57$\leq$Z$\leq$71). These elements can be produced 
in both the slow and rapid neutron-capture processes, but \citet{Truran1981} argued that in EMP stars these elements are predominantly 
due to r-processes because intermediate-mass AGB stars had no time to enrich the matter before the formation of the observed 
EMP stars.

This work is organized as follows. Section 2 describes the observations and data reduction; Sect. 3 summarizes the procedures of 
abundance determination, as well as the adopted final abundances for each element; Sect. 4 discusses the results in the 
context of r-process models; Sect. 5 summarizes our conclusions.

%__________________________________________________________________

\section {Observations}

CS 31082-001 was observed with the Space Telescope Imaging Spectrograph (STIS) in the near UV 
(program ID 9359; PI: R. Cayrel). STIS spectroscopic mode E230M combines an \'echelle grating with an NUV-MAMA 
detector to obtain spectra in the wavelength range 1575-3100 {\rm \AA}, at a resolution of R=30000. More information 
can be found in \citet{Barbuy2011}.

As a complement to these observations we used the mean of three UVES spectra centered at 340 nm, obtained in 2001 at the 
VLT on October 20, 21, and 22, in the framework of the ESO Large Program ``First Stars''. In the common range of 
wavelengths ($\rm300 nm < \lambda < 307 nm$), the co-added spectrum has a higher resolution (R=75000) than the HST spectrum, 
and a S/N of about 20 at 300 nm and 100 at 340 nm, and it was not used by \citet{Hill2002}. The reduction 
of these spectra was performed using the UVES context within MIDAS: bias subtraction, fit and subtraction of the interorder 
background from the object, and flat-field images. The wavelength calibration was performed on Th-Ar lamp frames and used to build a 
co-added spectrum. 

%__________________________________________________________________

\section{Abundance determination}

The present abundance determinations are based on the OSMARCS LTE atmospheric model \citep{Gustafsson2003,Gustafsson2008}, 
which use an updated version of the MARCS program \citep{Gustafsson1975,Plez1992,Asplund1997} to build 1D LTE plane-parallel models for cool stars. 
As done by \citet{Hill2002} and \citet{Barbuy2011}, we used a consistent approach based on the spectrum synthesis 
code Turbospectrum \citep{Alvarez1998}, which includes a full chemical equilibrium 
and Van der Waals collisional broadening 
by H, He, and H$_{2}$, following \citet{Anstee1995}, \citet{Barklem1997}, and \citet{Barklem1998}. 
The code also properly accounts for scattering in the continuum, an 
important effect in the UV \citep{Cayrel2004,Barbuy2011}. 
The stellar parameters are adopted from \citet{Hill2002}: T$_{\rm eff}$ = 4825 $\pm$ 50 K, log $g$ = 1.5 $\pm$ 0.3, 
[Fe/H] = -2.9 $\pm$ 0.1 (in LTE), and  $v_t$ = 1.8 $\pm$ 0.2 km s$^{-1}$. We also adopted the light element abundances determined by 
\citet{Hill2002}, \citet{Cayrel2004}, and \citet{Spite2005}.

The calculations used the Turbospectrum molecular line lists, described detailedly in \citet{Alvarez1998}, together with 
the atomic line lists from the VALD2 compilation \citep{Kupka1999}, unless updated oscillator strengths were available in the 
literature: 
Cr I from \citet{Sobeck2007}; 
Ge I from \citet{Biemont1999}; 
La II from \citet{Lawler2001a}; 
Ce II from \citet{Palmeri2000} and \citet{Lawler2009}; 
Nd II from \citet{Den Hartog2003}; 
Sm II from \citet{Lawler2006}; 
Eu I and Eu II from \citet{Lawler2001b} and \citet{Den Hartog2002}; 
Gd II from \citet{Den Hartog2006}; 
Tb II from \citet{Lawler2001c}; 
Dy II from \citet{Sneden2009}; 
Er II from \citet{Lawler2008}; 
Tm I and Tm II from \citet{Wickliffe1997} and \citet{Sneden2009}; 
Lu I from \citet{Fedchak2000}; 
Lu II from \citet{Quinet1999}; 
Hf II from \citet{Lawler2007}; 
Ta I from \citet{Fivet2006}; 
W II from \citet{Nilsson2008}.

Following the same procedure as in \citet{Barbuy2011}, we computed synthetic spectra for all lines of the elements of interest 
from our line list, with different enhancement factors, in order to identify useful lines. All lines were checked for proper 
intensities and possible blends, and lines with major and/or uncertain blends were rejected.
As discussed by \citet{Peterson2001}, modeling the UV region is difficult because of the crowding of lines at short wavelengths. 
Another extreme problem is the number of ``unknown'' lines (absent from the input line list), resulting in difficulties to match 
observations and spectral synthesis calculations, making it more difficult to normalize the UV continuum. 
Recently, \citet{Peterson2011} has reported determinations of the 
molybdenum abundances in five mildly to extremely metal-poor turnoff stars using near-UV spectra with a ``guessed identifications'' 
method of missing lines \citep[for the details on the procedure, see][]{Peterson2001}. 
In this work we preferred to reject those lines seriously affected by these effects.

\begin{table}
\caption{LTE abundances in CS~31082-001 as derived from previous works, from the present paper, and our adopted final abundances.}             % title of Table
\scalefont{0.75}
\label{abs}      % is used to refer this table in the text
\centering                          % used for centering table
\begin{tabular}{c c c c c c c c}        % centered columns (4 columns)
\hline\hline                 % inserts double horizontal lines
\hbox{El.} & \hbox{Z} & \hbox{A(X)} & \hbox{A(X)} & \hbox{A(X)} & \hbox{A(X)} & \hbox{A(X)} & \hbox{[X/Fe]}  \\
  &  & \hbox{(1)} & \hbox{(2)} & \hbox{(3)} & \hbox{This Work} & \hbox{\rm adopted} & \hbox{\rm adopted}  \\
\hline                        % inserts single horizontal line
\hbox{Ge} & 32 &  ---  &  ---  &  ---  & +0.10 & +0.10$\pm$0.21 & -0.55  \\
\hbox{Sr} & 38 & +0.72 &  ---  &  ---  &  ---  & +0.72$\pm$0.10 & 0.73 \\
\hbox{Y}  & 39 & -0.23 &  ---  &  ---  & -0.15 & -0.19$\pm$0.07 & 0.53 \\
\hbox{Zr} & 40 & +0.43 &  ---  &  ---  & +0.55 & +0.49$\pm$0.08 & 0.84 \\
\hbox{Nb} & 41 & -0.55 &  ---  &  ---  & -0.52 & -0.54$\pm$0.12 & 0.97 \\
\hbox{Mo} & 42 &  ---  &  ---  &  ---  & -0.11 & -0.11$\pm$0.13 & 0.90 \\
\hbox{Ru} & 44 & +0.36 &  ---  &  ---  & +0.36 & +0.36$\pm$0.12 & 1.45 \\
\hbox{Rh} & 45 & -0.42 &  ---  &  ---  & -0.42 & -0.42$\pm$0.12 & 1.39 \\
\hbox{Pd} & 46 & -0.05 &  ---  &  ---  & -0.09 & -0.09$\pm$0.07 & 1.18 \\
\hbox{Ag} & 47 & -0.81 &  ---  &  ---  & -0.84 & -0.84$\pm$0.21 & 1.15 \\
\hbox{Ba} & 56 & +0.40 &  ---  &  ---  &  ---  & +0.40$\pm$0.14 & 1.16 \\
\hbox{La} & 57 & -0.60 & -0.62 &  ---  &  ---  & -0.62$\pm$0.05 & 1.17 \\
\hbox{Ce} & 58 & -0.31 & -0.29 &  ---  & -0.31 & -0.29$\pm$0.05 & 1.03 \\
\hbox{Pr} & 59 & -0.86 & -0.79 &  ---  &  ---  & -0.79$\pm$0.05 & 1.38 \\
\hbox{Nd} & 60 & -0.13 & -0.15 &  ---  & -0.21 & -0.15$\pm$0.05 & 1.33 \\
\hbox{Sm} & 62 & -0.51 & -0.42 &  ---  & -0.42 & -0.42$\pm$0.05 & 1.51 \\
\hbox{Eu} & 63 & -0.76 & -0.72 &  ---  & -0.75 & -0.72$\pm$0.05 & 1.69 \\
\hbox{Gd} & 64 & -0.27 & -0.21 &  ---  & -0.29 & -0.21$\pm$0.05 & 1.61 \\
\hbox{Tb} & 65 & -1.26 & -1.01 &  ---  & -1.00 & -1.01$\pm$0.05 & 1.64 \\
\hbox{Dy} & 66 & -0.21 & -0.07 &  ---  & -0.12 & -0.07$\pm$0.05 & 1.73 \\
\hbox{Ho} & 67 &  ---  & -0.80 &  ---  &  ---  & -0.80$\pm$0.06 & 1.62 \\
\hbox{Er} & 68 & -0.27 & -0.30 &  ---  & -0.31 & -0.30$\pm$0.05 & 1.67 \\
\hbox{Tm} & 69 & -1.24 & -1.15 &  ---  & -1.18 & -1.15$\pm$0.05 & 1.64 \\
\hbox{Yb} & 70 &  ---  & -0.41 &  ---  &  ---  & -0.41$\pm$0.11 & 1.66 \\
\hbox{Lu} & 71 &  ---  &  ---  &  ---  & -1.08 & -1.08$\pm$0.13 & 1.73 \\
\hbox{Hf} & 72 & -0.59 & -0.72 &  ---  & -0.73 & -0.72$\pm$0.05 & 1.33 \\
\hbox{Ta} & 73 &  ---  &  ---  &  ---  & -1.60 & -1.60$\pm$0.23 & 1.47 \\
\hbox{W}  & 74 &  ---  &  ---  &  ---  & -0.90 & -0.90$\pm$0.24 & 0.92 \\
\hbox{Re} & 75 &  ---  &  ---  &  ---  & -0.21 & -0.21$\pm$0.21 & 2.45 \\
\hbox{Os} & 76 & +0.43 &  ---  & +0.18 &  ---  & +0.18$\pm$0.07 & 1.72 \\
\hbox{Ir} & 77 & +0.20 &  ---  & +0.20 &  ---  & +0.20$\pm$0.07 & 1.72 \\
\hbox{Pt} & 78 &  ---  &  ---  & +0.30 &  ---  & +0.30$\pm$0.23 & 1.46 \\
\hbox{Au} & 79 &  ---  &  ---  & -1.00 &  ---  & -1.00$\pm$0.34 & 0.89 \\
\hbox{Pb} & 82 &  ---  &  ---  & -0.65 &  ---  & -0.65$\pm$0.19 & 0.25 \\
\hbox{Bi} & 83 &  ---  &  ---  & -0.40 &  ---  & -0.40$\pm$0.33 & 1.83 \\
\hbox{Th} & 90 & -0.98 &  ---  &  ---  &  ---  & -0.98$\pm$0.13 & 1.84 \\
\hbox{U}  & 92 & -1.92 &  ---  &  ---  &  ---  & -1.92$\pm$0.17 & 1.68 \\
\hline                                   %inserts single line
\end{tabular}
\tablebib{(1) \citet{Hill2002}, (2) \citet{Sneden2009}, (3) \citet{Barbuy2011}.}
\end{table}

\subsection{Final abundances}

The selected lines and individual abundances are available in Table \ref{lines_total}.
The mean abundances A(X)\footnote{
We adopt the notation A(X) = log $\epsilon$(X) = log n(X)/n(H) + 12, with n = number density of atoms.} 
for 23 neutron capture elements are given in Table \ref{abs} (column 6) and compared to previous 
measurements \citep{Hill2002,Sneden2009}. We were able to derive the abundances of six newly studied 
elements: germanium (Ge, Z=32), molybdenum (Mo, Z=42), lutetium (Lu, Z=71), tantalum (Ta, Z=73), tungsten (W, Z=74), 
and rhenium (Re, Z=75). The general agreement is discussed in section 3.3. 
We also investigated the elements in the region between germanium and strontium, as well the elements between the first and the second peaks of the r-process, 
but no useful line was found. 

As noted above, many elements of the first and second peaks of the r-process in 
these metal-poor stars are observable from the ground, 
and several authors have presented analyses of them, so for each element, we compared the new abundance with the previous data in 
order to adopt a final value. The results are shown in the columns 7 and 8 of Table \ref{abs}.

For comparison, the solar abundances from different authors in the literature \citep{Anders1989,Grevesse1998,Sneden2008,Asplund2009,Lodders2009,Caffau2011} 
are listed in Table \ref{sol} for all elements of interest. In this work 
we adopted the values from \citet{Lodders2009} and \citet{Caffau2011}. 
We also present the Solar System r- and s-process deconvolution of \citet{Simmerer2004},. 
 and use the fractions of r-process from the deconvolution of 
\citet{Simmerer2004} with the adopted solar system abundances.

\begin{table}
\caption{Solar {\it r}- and {\it s}-process fractions \citep{Simmerer2004} and total solar abundances elements. 
Adopted abundances are marked in boldface. $^*$: Meteoritic abundances.}             % title of Table
\scalefont{0.75}
\label{sol}      % is used to refer this table in the text
\centering                          % used for centering table
\begin{tabular}{c c c c c c c c c}        % centered columns (4 columns)
\hline\hline                 % inserts double horizontal lines
\hbox{El.} & \hbox{Z} & \multicolumn{2}{c}{Fraction} & \multicolumn{5}{c}{\hbox{A(X)$_{\odot}$}}  \\
\cline{3-4} \cline{5-9}   \\
 &  & r & s & (1) & (2)  & (3) & (4) & (5) \\
\hline                        % inserts single horizontal line
\hbox{Fe} & 26 &  --   &  --   & 7.67  &  7.50  &  7.50  &  7.45  & \bf{7.52} \\
\hbox{Ge} & 32 & 0.516 & 0.484 & 3.41  &  3.41  &  3.65  &  \bf{3.58}  & -- \\
\hbox{Sr} & 38 & 0.110 & 0.890 & 2.90  &  2.97  &  2.87  &  \bf{2.92}  & -- \\
\hbox{Y}  & 39 & 0.281 & 0.719 & 2.24  &  2.24  &  2.21  &  \bf{2.21}  & -- \\
\hbox{Zr} & 40 & 0.191 & 0.809 & 2.60  &  2.60  &  2.58  &  \bf{2.58}  & -- \\
\hbox{Nb} & 41 & 0.324 & 0.676 & 1.42  &  1.42  &  1.46  &  \bf{1.42}  & -- \\
\hbox{Mo} & 42 & 0.323 & 0.677 & 1.92  &  1.92  &  1.88  &  \bf{1.92}  & -- \\
\hbox{Ru} & 44 & 0.610 & 0.390 & 1.84  &  1.84  &  1.75  &  \bf{1.84}  & -- \\
\hbox{Rh} & 45 & 0.839 & 0.161 & 1.12  &  1.12  &  0.91  &  \bf{1.12}  & -- \\
\hbox{Pd} & 46 & 0.555 & 0.445 & 1.69  &  1.69  &  1.57  &  \bf{1.66}  & -- \\
\hbox{Ag} & 47 & 0.788 & 0.212 & 1.24* &  1.24* &  0.94  &  \bf{0.94}  & -- \\
\hbox{Sn} & 50 & 0.225 & 0.775 & 2.0   &  2.0   &  2.04  &  \bf{2.00}  & -- \\
\hbox{Ba} & 56 & 0.147 & 0.853 & 2.13  &  2.13  &  2.18  &  \bf{2.18}  & -- \\
\hbox{La} & 57 & 0.246 & 0.754 & 1.22  &  1.17  &  1.10  &  \bf{1.14}  & -- \\
\hbox{Ce} & 58 & 0.186 & 0.814 & 1.55  &  1.58  &  1.58  &  \bf{1.61}  & -- \\
\hbox{Pr} & 59 & 0.508 & 0.492 & 0.71  &  0.71  &  0.72  &  \bf{0.76}  & -- \\
\hbox{Nd} & 60 & 0.421 & 0.579 & 1.50  &  1.50  &  1.42  &  \bf{1.45}  & -- \\
\hbox{Sm} & 62 & 0.669 & 0.331 & 1.00  &  1.01  &  0.96  &  \bf{1.00}  & -- \\
\hbox{Eu} & 63 & 0.973 & 0.027 & 0.51  &  0.51  &  0.52  &  0.52  & \bf{0.52} \\
\hbox{Gd} & 64 & 0.819 & 0.181 & 1.12  &  1.12  &  1.07  &  \bf{1.11}  & -- \\
\hbox{Tb} & 65 & 0.933 & 0.067 & 0.33* &  0.35* &  0.30  &  \bf{0.28}  & -- \\
\hbox{Dy} & 66 & 0.879 & 0.121 & 1.10  &  1.14  &  1.10  &  \bf{1.13}  & -- \\
\hbox{Ho} & 67 & 0.936 & 0.064 & 0.26  &  0.26  &  0.48  &  \bf{0.51}  & -- \\
\hbox{Er} & 68 & 0.832 & 0.168 & 0.93  &  0.93  &  0.92  &  \bf{0.96}  & -- \\
\hbox{Tm} & 69 & 0.829 & 0.171 & 0.13* &  0.15* &  0.10  &  \bf{0.14}  & -- \\
\hbox{Yb} & 70 & 0.682 & 0.318 & 1.08  &  1.08  &  0.84  &  \bf{0.86}  & -- \\
\hbox{Lu} & 71 & 0.796 & 0.204 & 0.12* &  0.06  &  0.10  &  \bf{0.12}  & -- \\
\hbox{Hf} & 72 & 0.510 & 0.490 & 0.88  &  0.88  &  0.85  &  0.88  & \bf{0.87} \\
\hbox{Ta} & 73 & 0.588 & 0.412 & 0.13* & -0.13* & -0.12* & \bf{-0.14}* & -- \\
\hbox{W}  & 74 & 0.462 & 0.538 & 0.68* &  0.69* &  0.85  &  \bf{1.11}  & -- \\
\hbox{Re} & 75 & 0.911 & 0.089 & 0.27* &  0.28* &  0.26* &  \bf{0.28}* & -- \\
\hbox{Os} & 76 & 0.916 & 0.084 & 1.45  &  1.45  &  1.40  &  1.45  & \bf{1.36} \\
\hbox{Ir} & 77 & 0.988 & 0.012 & 1.35  &  1.35  &  1.38  &  \bf{1.38}  & -- \\
\hbox{Pt} & 78 & 0.949 & 0.051 & 1.8   &  1.8   &  1.62* &  \bf{1.74}  & -- \\
\hbox{Au} & 79 & 0.944 & 0.056 & 1.01  &  1.01  &  0.92  &  \bf{1.01}  & -- \\
\hbox{Pb} & 82 & 0.214 & 0.786 & 1.85  &  1.95  &  1.75  &  \bf{2.00}  & -- \\
\hbox{Bi} & 83 & 0.647 & 0.353 & 0.71* &  0.71* &  0.65* &  \bf{0.67}* & -- \\
\hbox{Th} & 90 & 1.000 & 0.000 & 0.12  &  0.08* &  0.02  & <0.08 & \bf{0.08} \\
\hbox{U}  & 92 & 1.000 & 0.000 & <-0.47 &  <-0.47 & -0.54* &  \bf{<-0.47} & -- \\
\hline                                   %inserts single line
\end{tabular}
\tablebib{(1): \citet{Anders1989}; (2): \citet{Grevesse1998}; (3): \citet{Asplund2009}; (4): \citet{Lodders2009}; (5): \citet{Caffau2011}.}
\end{table}

\subsection{Uncertainties on the derived abundances}

As discussed by \citet{Cayrel2004}, for a given stellar temperature, the ionization equilibrium provides an estimate of the 
stellar gravity with an internal accuracy of about 0.1 dex in log $g$, and the microturbulence velocity $v_{t}$ can be constrained 
within 0.2 km s$^{-1}$, making of the temperature the largest source of uncertainties in the abundance determination. In fact, 
the authors estimate that the total error on the adopted temperatures is on the order of 100 K, higher than the previous estimation 
found by \citet{Hill2002}. 

We estimate the abundance uncertainties arising from each of these three sources independently. The results are shown in 
the Table \ref{erro_model} (columns 3 to 5), where the models B, C, and D are compared with our nominal model labeled A. 

Since the stellar parameters are not independent of each other, the total error budget is not the quadratic sum of the various sources of 
uncertainties, but it does contain significant covariance terms. The solution was to create a new atmospheric model with a
temperature higher by 100 K, thereby determining the corresponding surface gravity by requiring that the Fe derived from Fe I and Fe II lines 
be identical, and the microturbulent velocity requiring that the abundance derived for individual Fe I lines to be independent 
of the equivalent width of the line. The model E is the result of this method, with $T_{\rm eff}$ = 4925 K, $\log g$ = 1.8 dex, and 
$v_{\rm t}$ = 1.8 km s$^{-1}$, and Table \ref{erro_model} (column 6) shows abundance uncertainties arising from stellar 
parameters.

Observational errors were estimated using the standard deviation of the abundances from the individual lines for each element, 
and they must take the uncertainties in defining the continuum, fitting the line profiles, and in the oscillator 
strengths  into account. 
For elements presenting only three useful lines, we conservatively adopted the typical observational error obtained
 for molybdenum as representative of these cases, 
and when the number of lines for a given element is only two or less we adopted 0.2 dex, as described in section 3.3.1.

Finally, we examined the adopted resolution for the synthetic spectrum calculation as another possible source of uncertainties. 
\citet{Barbuy2011} discuss in detail this value (R=30 800) which includes the effect of the instrumental profile, the macroturbulence and the rotational 
velocities of the stars. Analyzing the lines again, we checked that a change of 8 $\%$ in
 R induces a variation in the abundances of at most 0.05 dex, 
which we take into account for all elements in the final error.

\begin{table}
\caption{Abundance uncertainties linked to stellar parameters.}             % title of Table
\label{erro_model}      % is used to refer this table in the text
\centering                          % used for centering table
\begin{tabular}{c c c c c}        % centered columns (4 columns)
\hline\hline                 % inserts double horizontal lines
\multicolumn{5}{c}{\hbox{A: $T_{eff}$ = 4825, log $g$ = 1.5 dex, $v_{t}$ = 1.8 km s$^{-1}$}}\\ 
\multicolumn{5}{c}{\hbox{B: $T_{eff}$ = 4825, log $g$ = 1.4 dex, $v_{t}$ = 1.8 km s$^{-1}$}}\\ 
\multicolumn{5}{c}{\hbox{C: $T_{eff}$ = 4825, log $g$ = 1.5 dex, $v_{t}$ = 1.6 km s$^{-1}$}}\\ 
\multicolumn{5}{c}{\hbox{D: $T_{eff}$ = 4925, log $g$ = 1.5 dex, $v_{t}$ = 1.8 km s$^{-1}$}}\\ 
\multicolumn{5}{c}{\hbox{E: $T_{eff}$ = 4925, log $g$ = 1.8 dex, $v_{t}$ = 1.8 km s$^{-1}$}}\\ 
\hline\hline                 % inserts double horizontal lines
\hbox{El.} & \hbox{$\Delta_{B-A}$} & \hbox{$\Delta_{C-A}$} & \hbox{$\Delta_{D-A}$} & \hbox{$\Delta_{E-A}$} \\
\hline                        % inserts single horizontal line
\hbox{[Fe/H]}     & -0.01 & +0.04 & +0.06 & +0.10 \\
\hbox{[Ge I/Fe]}  & +0.01 & -0.03 & +0.09 & +0.05 \\
\hbox{[Y II/Fe]}  & -0.01 & +0.12 & +0.03 & +0.05 \\
\hbox{[Zr II/Fe]} & -0.01 & +0.06 & +0.02 & +0.06 \\
\hbox{[Nb II/Fe]} & -0.02 & -0.02 & +0.03 & +0.09 \\
\hbox{[Mo II/Fe]} & -0.02 & -0.02 & +0.01 & +0.07 \\
\hbox{[Ru I/Fe]}  & +0.02 & -0.03 & +0.10 & +0.04 \\
\hbox{[Rh I/Fe]}  & +0.02 & -0.04 & +0.09 & +0.04 \\
\hbox{[Pd I/Fe]}  & +0.02 & -0.02 & +0.09 & +0.04 \\
\hbox{[Ag I/Fe]}  & +0.02 & -0.03 & +0.10 & +0.04 \\
\hbox{[Ce II/Fe]} & -0.02 & -0.03 & +0.03 & +0.08 \\
\hbox{[Nd II/Fe]} & -0.02 & -0.03 & +0.03 & +0.08 \\
\hbox{[Sm II/Fe]} & -0.02 & -0.02 & +0.03 & +0.08 \\
\hbox{[Eu II/Fe]} & -0.01 & +0.07 & +0.03 & +0.07 \\
\hbox{[Gd II/Fe]} & -0.02 & +0.01 & +0.03 & +0.07 \\
\hbox{[Tb II/Fe]} & -0.02 & -0.02 & +0.03 & +0.08 \\
\hbox{[Dy II/Fe]} & -0.01 & +0.07 & +0.03 & +0.06 \\
\hbox{[Er II/Fe]} & -0.02 & +0.01 & +0.02 & +0.07 \\
\hbox{[Tm II/Fe]} & -0.02 & -0.01 & +0.02 & +0.07 \\
\hbox{[Lu II/Fe]} & -0.02 & -0.03 & +0.01 & +0.07 \\
\hbox{[Hf II/Fe]} & -0.02 & -0.01 & +0.03 & +0.08 \\
\hbox{[Ta II/Fe]}  & -0.03 & -0.03 & +0.04 & +0.11 \\
\hbox{[W II/Fe]}  & -0.03 & -0.02 & +0.05 & +0.12 \\
\hbox{[Re II/Fe]} & -0.03 & -0.02 & -0.01 & +0.06 \\
\hline                                   %inserts single line
\end{tabular}
\end{table}

\subsection{Elements of the first peak}

\subsubsection{Germanium (Z=32)}

\begin{figure}
%\resizebox{\hsize}{!}{\includegraphics[angle=0,scale=.5]{Zr.pdf}}
\centering
\resizebox{70mm}{!}{\includegraphics{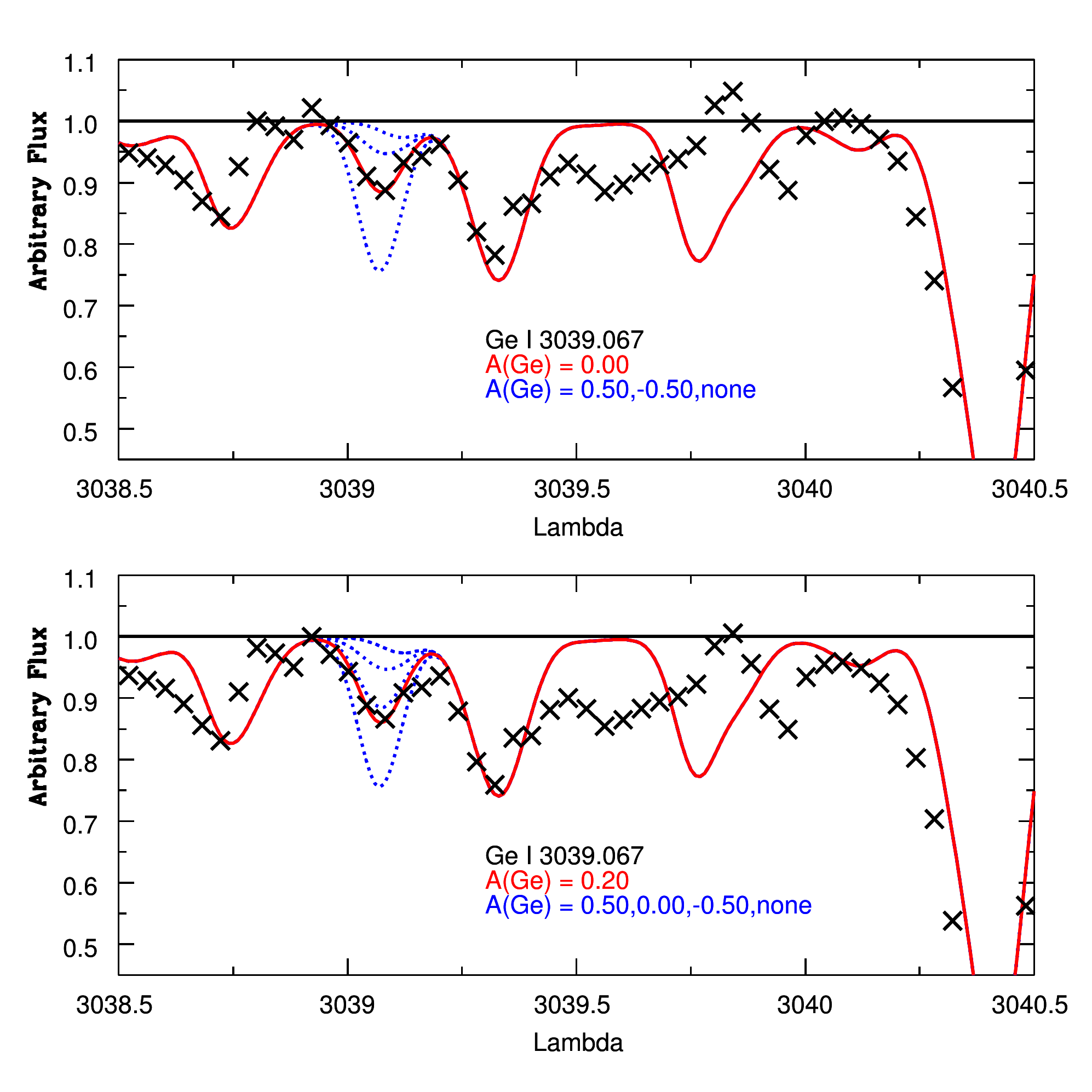}}
\caption{Fits of the observed Ge I 3039.067 {\AA} line in CS 31082-001 for two different placements of the continuum. 
Crosses: observations. Dotted lines: synthetic spectra computed for the abundances indicated in the figure. 
Solid line: synthetic spectrum computed with the best abundance, also indicated in the figure.}
\label{Ge}
\end{figure}

The Ge I 3039.067 {\AA} line is the main abundance indicator for this element. Using the \citet{Biemont1999} gf-value we 
were able to determine an abundance of A(Ge)=0.10$\pm$0.21 dex, a first detection of this element in CS 31082-001. 
Since we only used one line, we checked the influence of the placement of the continuum on the derived abundance, as shown in Fig. \ref{Ge}, 
and we assumed an observational error of $\sigma$=0.2 dex as a good estimation of this value. We also used this observation uncertainty with other elements 
which present only two lines or a single one. In fact, since such lines have a better definition of the continuum than the Ge line, we assume this 0.2 dex value as an upper limit, 
in particular with the UVES spectrum. 

Our result of [Ge/Fe]=-0.55 agrees with those found by \citet{Cowan2002,Cowan2005} in metal-poor galactic halo stars, showing that 
germanium is strongly depleted compared to the solar abundance ratio, even in r-rich metal-poor stars.

\subsubsection{Yttrium (Z=39)}

Using 13 new Y II lines with other two lines already used in \citet{Hill2002}, we are able to determine an abundance of 
A(Y)=-0.15$\pm$0.07 dex for this element, in agreement with A(Y)=-0.23$\pm$0.12 dex found previously. In fact, both of the 
lines in common between \citet{Hill2002} and the present work show compatible results, leading us to adopt the 
average A(Y)=-0.19$\pm$0.07 dex ([Y/Fe]=0.53) as the final abundance.

Recently, \citet{Hansen2012} analyzed a sample of metal-poor stars, including CS 31082-001. They also adopted MARCS models,
but for CS 31082-001, a slightly different set of atmospheric parameters (T$_{eff}$=4925 K, log$g$=1.51 [cgs], V$_{t}$=1.4, [Fe/H]=-2.81). 
For the spectrum synthesis they adopted MOOG \citep[][version 2009 including treatment of scattering]{Sneden1973}. 
The difference between their [X/Fe] results and our adopted values is in general lower than 0.1, but for
 yttrium they found [Y/Fe]=0.82,  which is 0.29 higher than our adopted value. 

We decided to check this comparison in a homogeneous way, using the code Turbospectrum to determine the Y abundance in CS 31082-001 using the lines from \citet{Hansen2012} 
with our set of atmospheric parameters as well as with their values.

We find [Y/Fe]=0.47 with our original atmospheric parameters, while the atmospheric parameters of \citet{Hansen2012} give the relative abundance [Y/Fe]=0.71, and the difference 
0.24 is very close to the previous one. However, we found a stronger correlation between the individual abundances and the equivalent widths with the atmospheric parameters 
adopted by \citet{Hansen2012}, suggesting that their microturbulence velocity is underestimated. In fact, when
only changing the $v_{t}$ in their set of atmospheric parameters 
to our microturbulence velocity we found [Y/Fe]=0.47, in agreement with our adopted value, confirming the problem with their $v_{t}$. 
Yttrium is particularly sensitive to this effect because several Y lines are relatively strong.

\subsubsection{Zirconium (Z=40)}

\begin{figure}
%\resizebox{\hsize}{!}{\includegraphics[angle=0,scale=.5]{Zr.pdf}}
\centering
\resizebox{70mm}{!}{\includegraphics{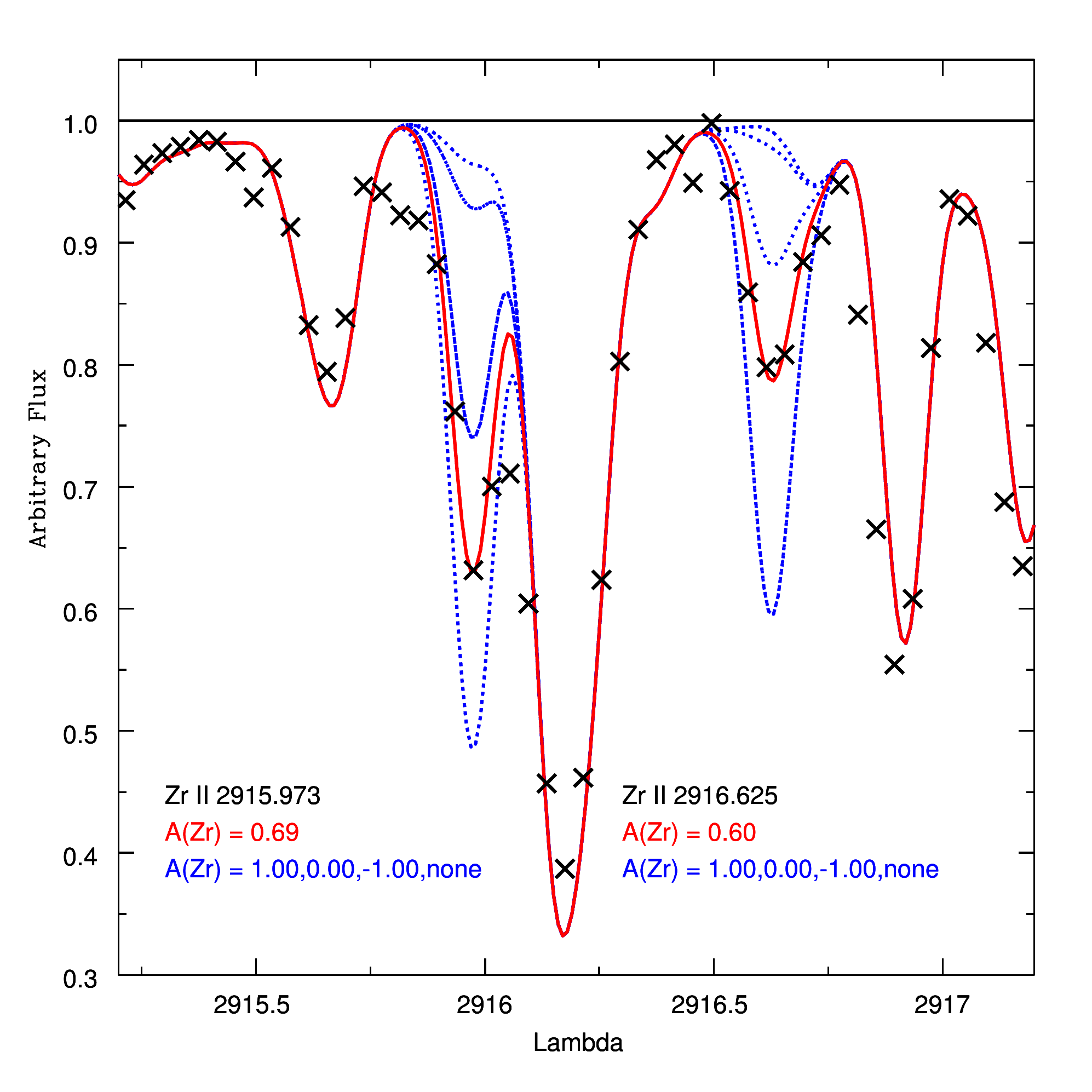}}
\caption{Fits of the observed Zr II 2915.973 {\AA} and Zr II 2916.625 {\AA} lines in CS 31082-001. Symbols as in Fig. 1.}
\label{Zr}
\end{figure}

More than 25 Zr II profiles were checked in STIS spetrum but we decided to keep the 12 best lines, and together with more 
46 UVES new useful lines, we find a final abundance A(Zr)=+0.55$\pm$0.08 dex for the zirconium, in agreement with the 
value A(Zr)=+0.43$\pm$0.14 dex from \citet{Hill2002}. Figure \ref{Zr} shows two fits of lines used in this work. 
The line Zr II 2758.792 {\AA} gives an abundance of A(Zr)=-0.07 dex, 
despite a good fit, suggesting there is a problem 
with the gf-value of the transition and leading us to exclude this line from the average. The same applies to Zr II 3556.585 {\AA} 
line, which gives A(Zr)=0.00 dex. 
We are using a larger set of lines than in previous work, and the abundances of the transitions used by \citet{Hill2002} can be considered as a 
subset of our data. Finally, the line Zr II 3030.915 {\AA} was used with both spectra STIS and UVES and we found compatible 
results. All this evidence suggests that the average A(Zr)=+0.49$\pm$0.08 dex ([Zr/Fe]=0.84) is the best choice for 
the adopted final abundance. 

\subsubsection{Niobium (Z=41)}

\begin{figure}
%\resizebox{\hsize}{!}{\includegraphics[angle=0]{Nb.pdf}}
\centering
\resizebox{70mm}{!}{\includegraphics{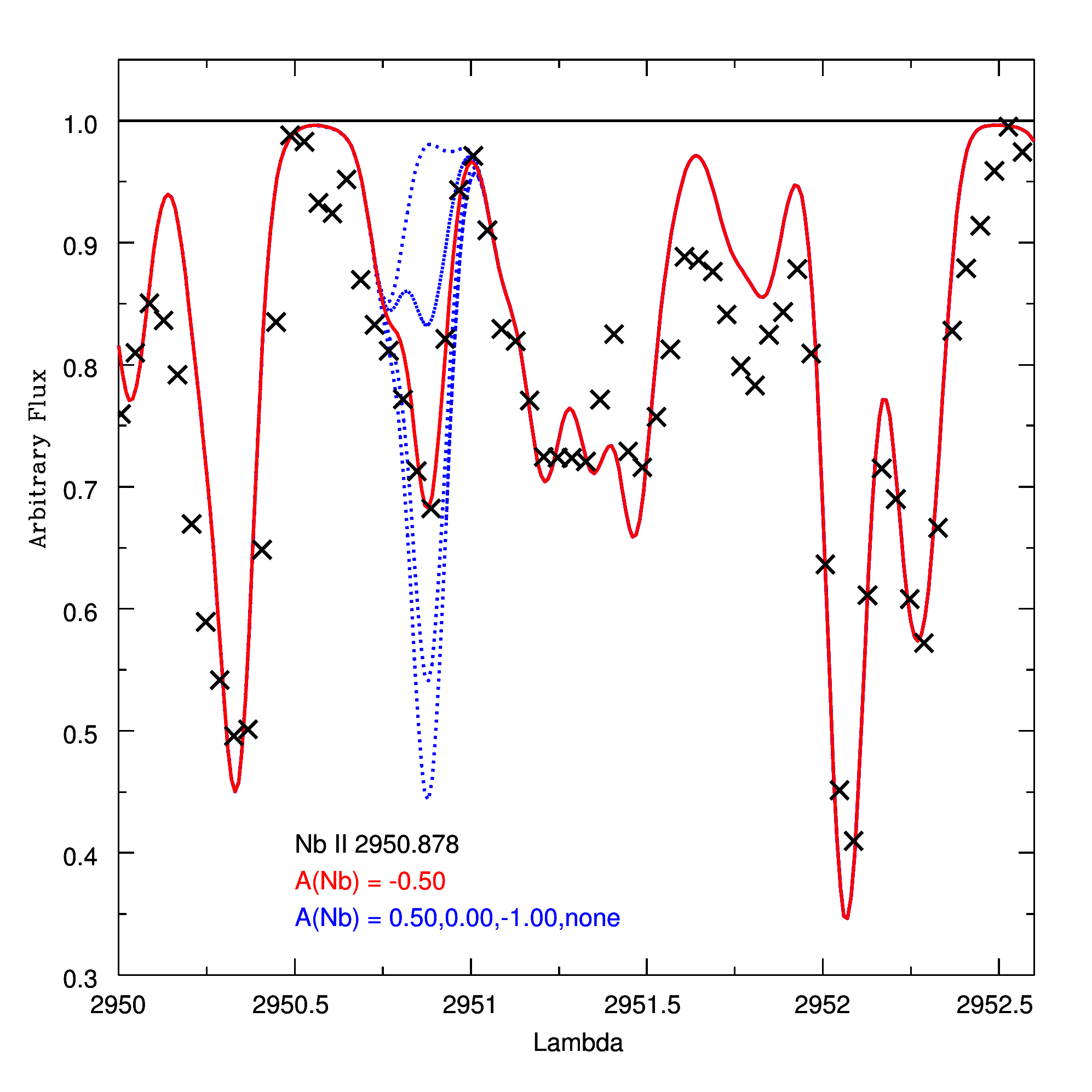}}
\caption{Fit of the observed Nb II 2950.878 {\AA} line in CS 31082-001. Symbols as in Fig. 1.}
\label{Nb}
\end{figure}

While only one line was used in \citet{Hill2002} to derive an abundance of A(Nb)=-0.55$\pm$0.20 dex, we were able to find 
nine useful Nb II lines from an initial set with more than 70 lines, 
giving an average abundance of A(Nb)=-0.52$\pm$0.11 dex, 
in good agreement with the previous value. In Fig. \ref{Nb} we show an example of fit
to a Nb II line. 

It was possible to again use both of the spectra with the same line Nb II 3028.433 {\AA}, and we found  agreement between the 
results. It makes the average a good indicator for the abundance of niobium, and we adopted 
A(Nb)=-0.54$\pm$0.12 dex ([Nb/Fe]=0.97) as the final value. 

\subsubsection{Molybdenum (Z=42)}

\begin{figure}
%\resizebox{\hsize}{!}{\includegraphics[angle=0]{Mo1.pdf}}
\centering
\resizebox{70mm}{!}{
\includegraphics{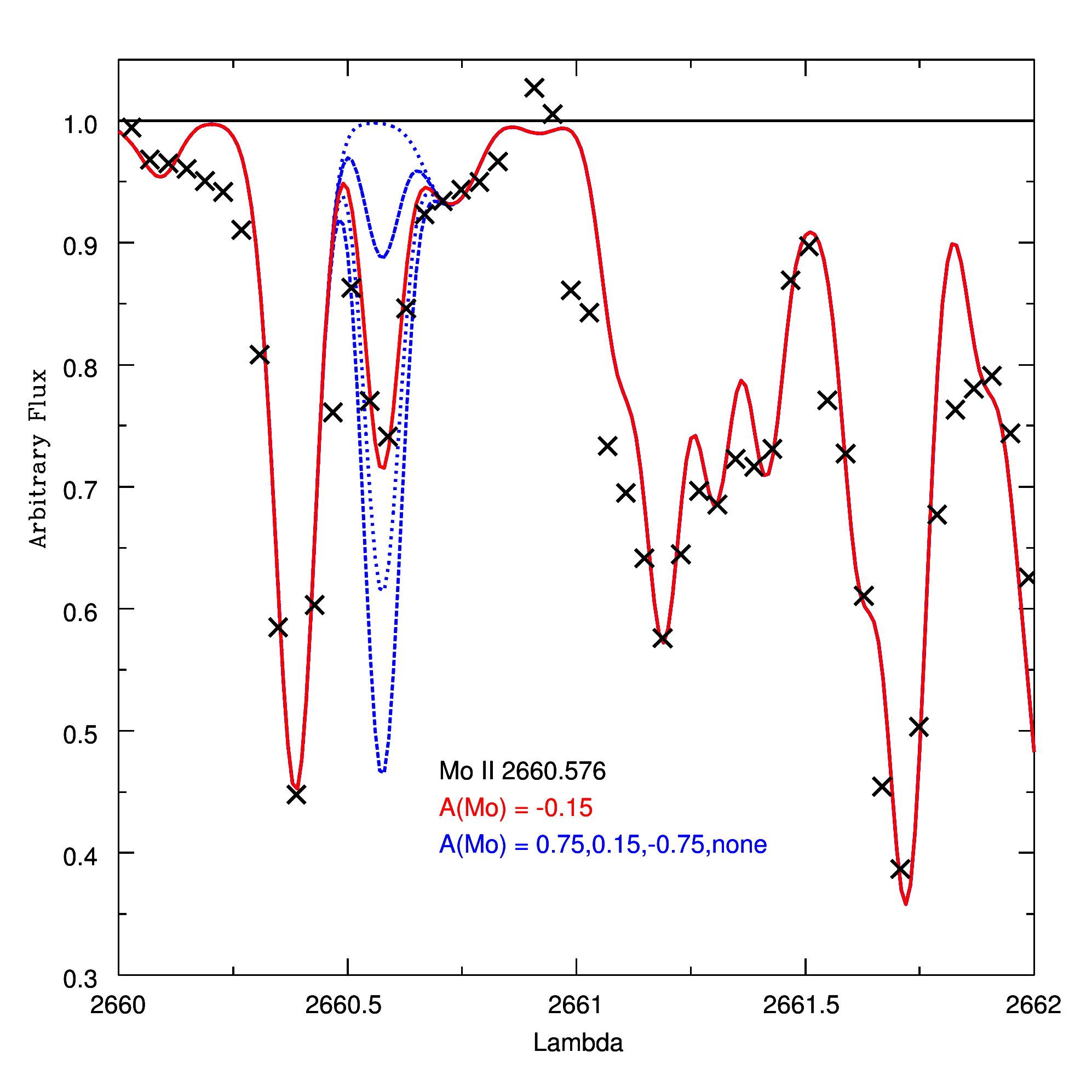}
}
\caption{Fit of the observed Mo II 2660.576 {\AA} line in CS 31082-001. Symbols as in Fig. 1.}
\label{Mo1}
\end{figure}

\begin{figure}
%\resizebox{\hsize}{!}{\includegraphics[angle=0]{Mo2.pdf}}
\centering
\resizebox{70mm}{!}{
\includegraphics{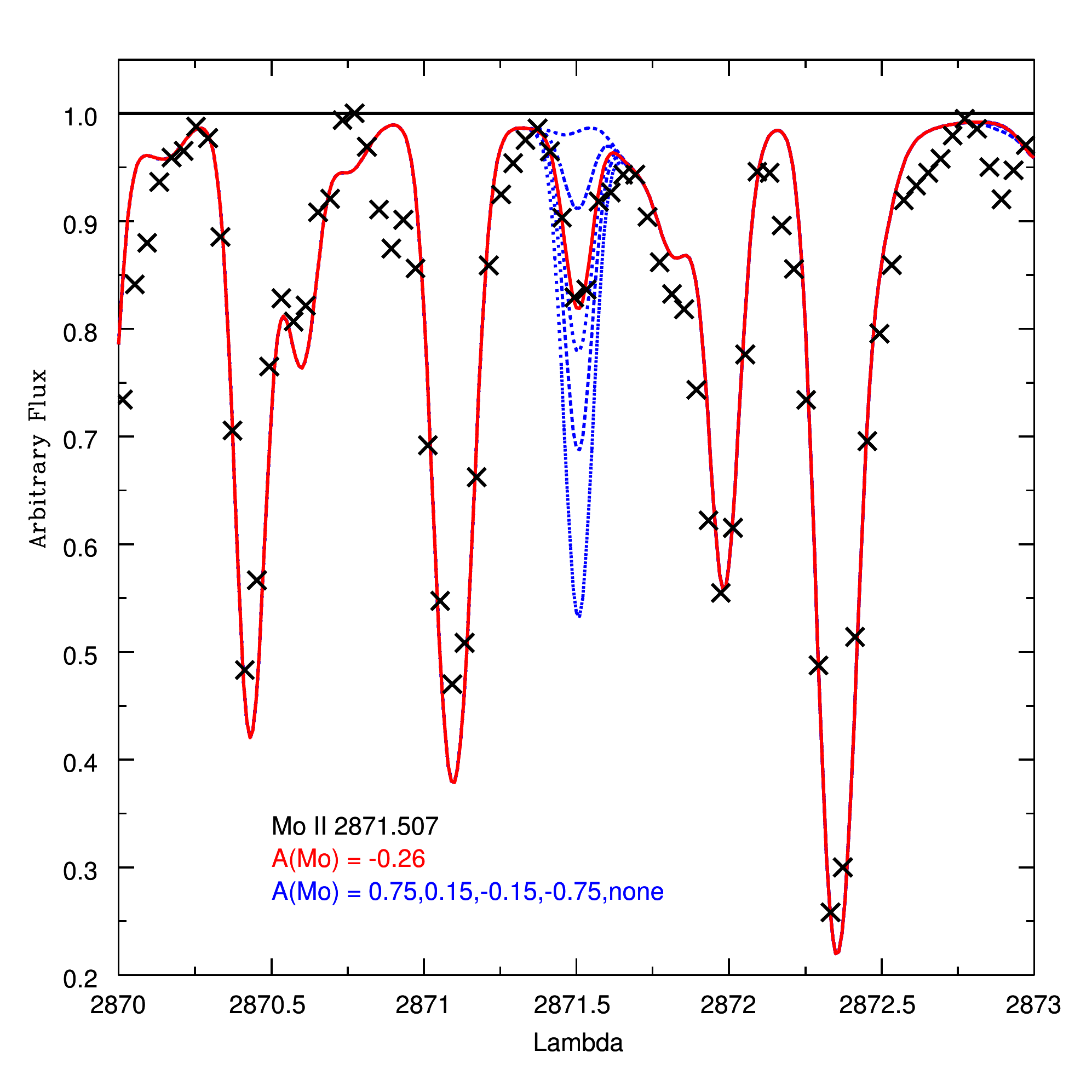}}
\caption{Fit of the observed Mo II 2871.507 {\AA} line in CS 31082-001. Symbols as in Fig. 1.}
\label{Mo2}
\end{figure}

\begin{figure}
\centering
\resizebox{70mm}{!}{\includegraphics{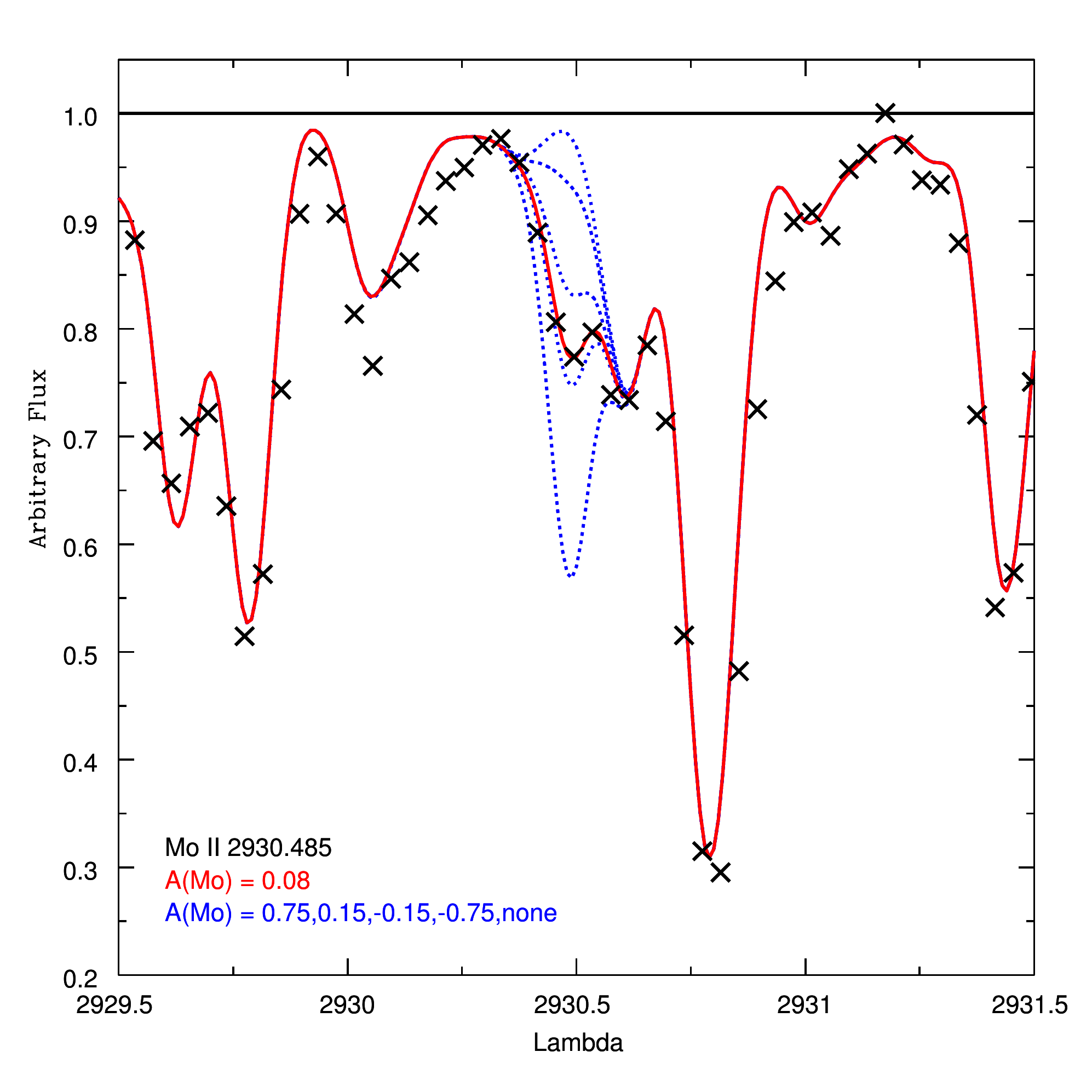}}
\caption{Fit of the observed Mo II 2930.485 {\AA} line in CS 31082-001. Symbols as in Fig. 1.}
\label{Mo3}
\end{figure}

After checking almost 50 Mo lines, the spectra presented three Mo II useful lines, giving A(Mo)=-0.11$\pm$0.13 dex ([Mo/Fe]=0.90). 
Despite all the 
previous analysis done, this is the first published value of A(Mo) in this star. The lines 2660.576 {\AA}, 2871.507 {\AA}, and 
2930.485 {\AA} used here are shown in Figs. \ref{Mo1}, \ref{Mo2}, and \ref{Mo3}. 

\begin{figure}
%\resizebox{\hsize}{!}{\includegraphics[angle=0]{Eu.pdf}}
\centering
\resizebox{70mm}{!}{\includegraphics{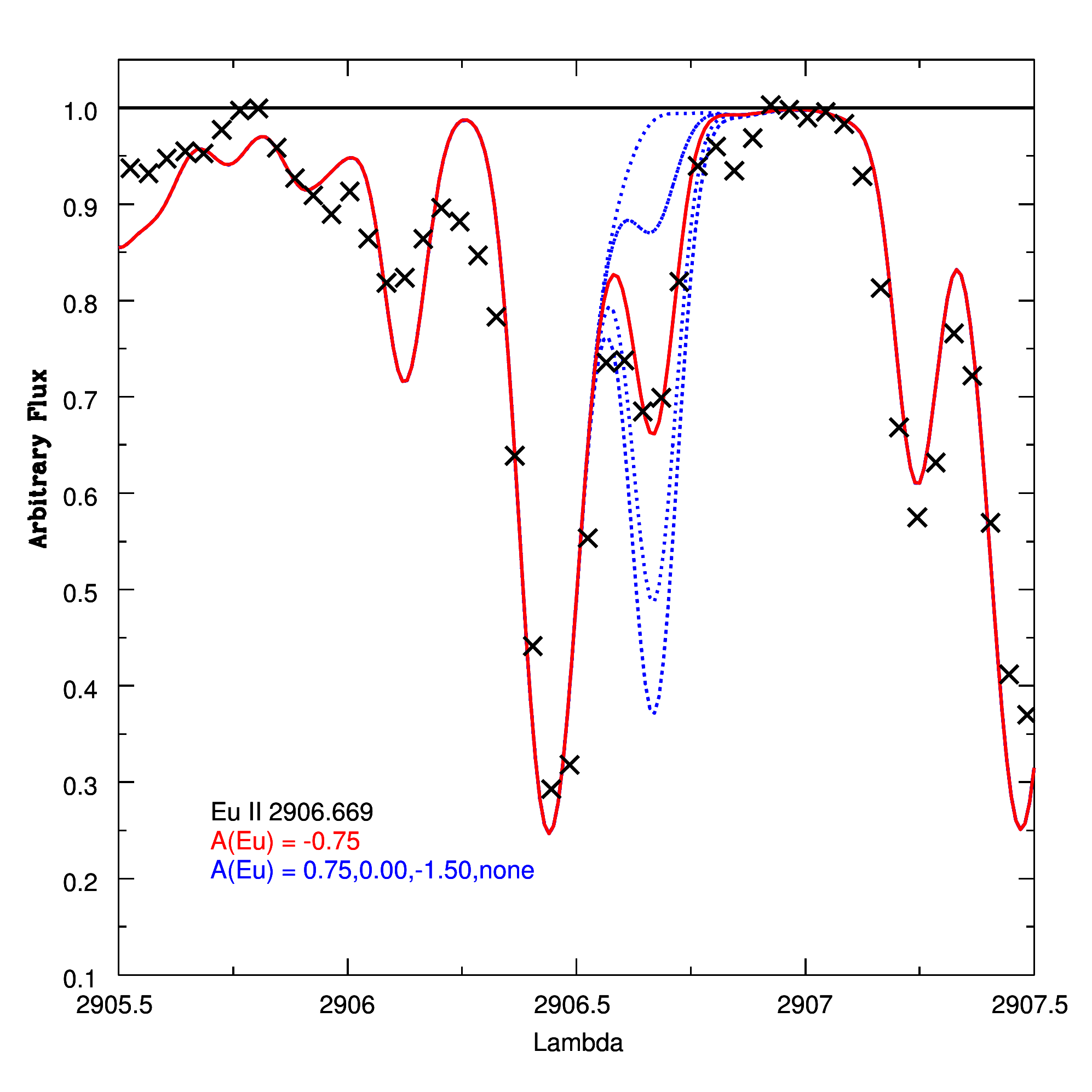}}
\caption{Fit of the observed Eu II 2906.669 {\AA} line in CS 31082-001. Symbols as in Fig. 1.}
\label{Eu}
\end{figure}

\begin{figure}
%\resizebox{\hsize}{!}{\includegraphics[angle=0]{Er.pdf}}
\centering
\resizebox{70mm}{!}{\includegraphics{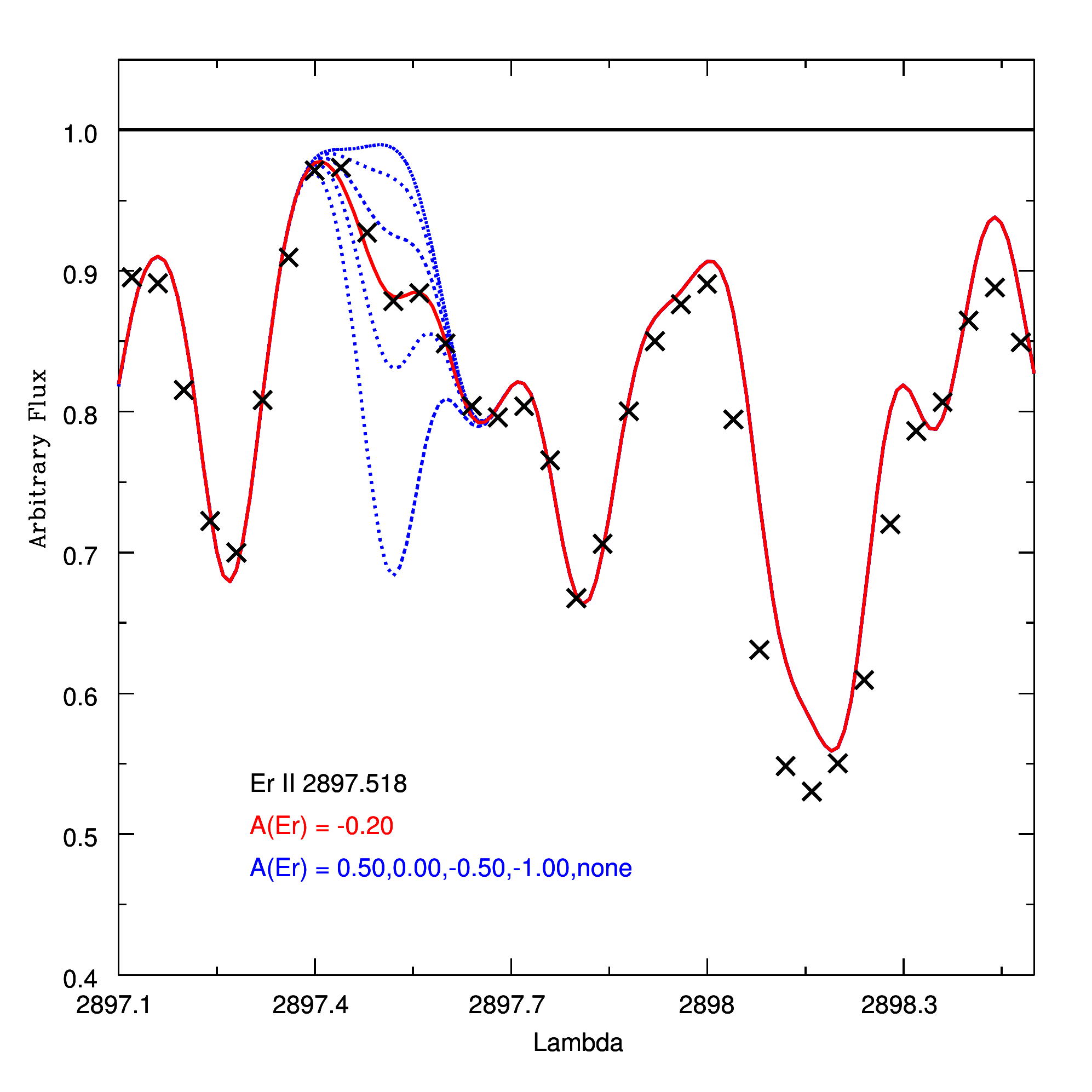}}
\caption{Fit of the observed Er II 2897.518 {\AA} line in CS 31082-001. Symbols as in Fig. 1.}
\label{Er}
\end{figure}

\begin{figure}
%\resizebox{\hsize}{!}{\includegraphics[angle=0]{Lu1.pdf}}
\centering
\resizebox{70mm}{!}{\includegraphics{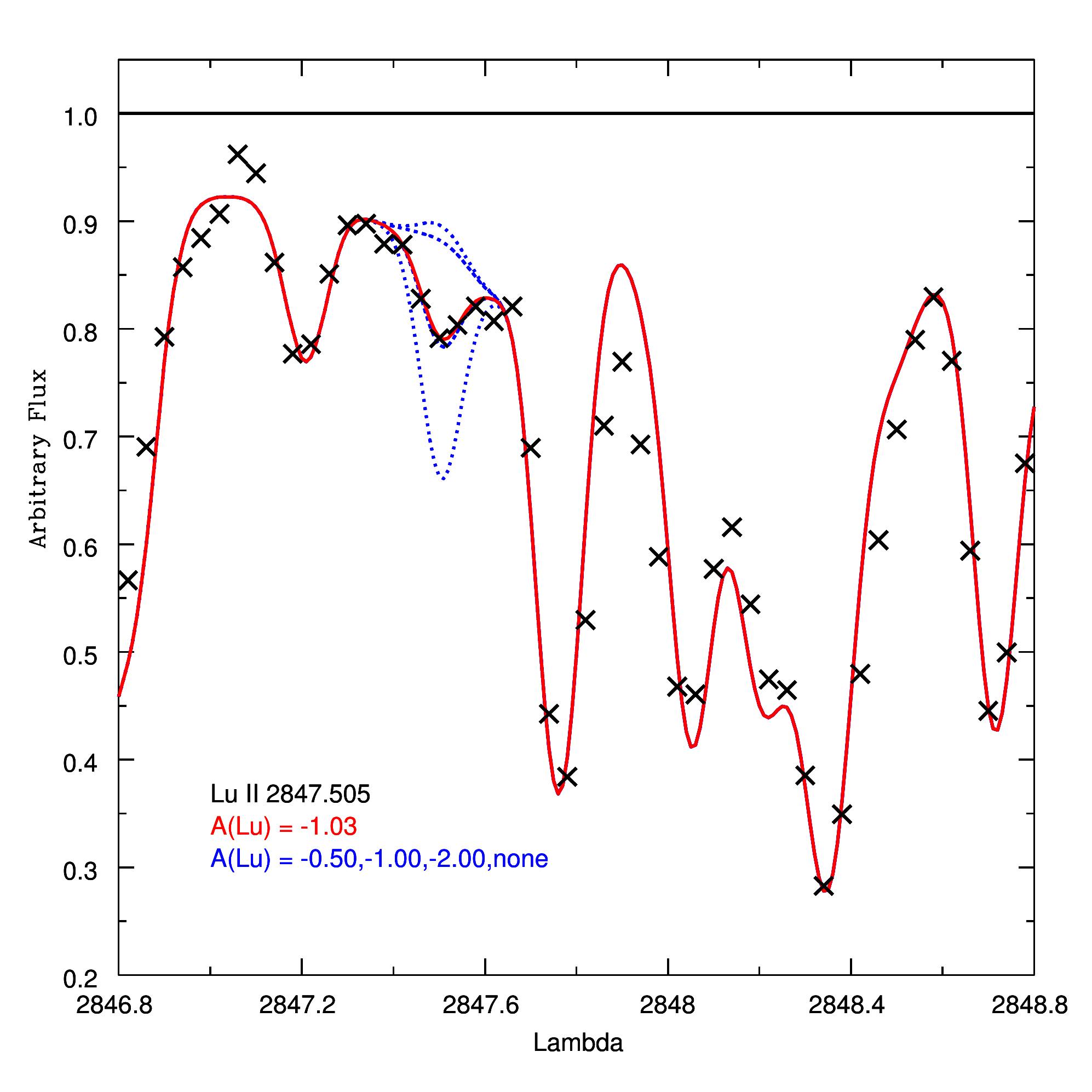}}
\caption{Fit of the observed Lu II 2847.505 {\AA} line in CS 31082-001. Symbols as in Fig. 1.}
\label{Lu1}
\end{figure}

\begin{figure}
\centering
\resizebox{70mm}{!}{\includegraphics{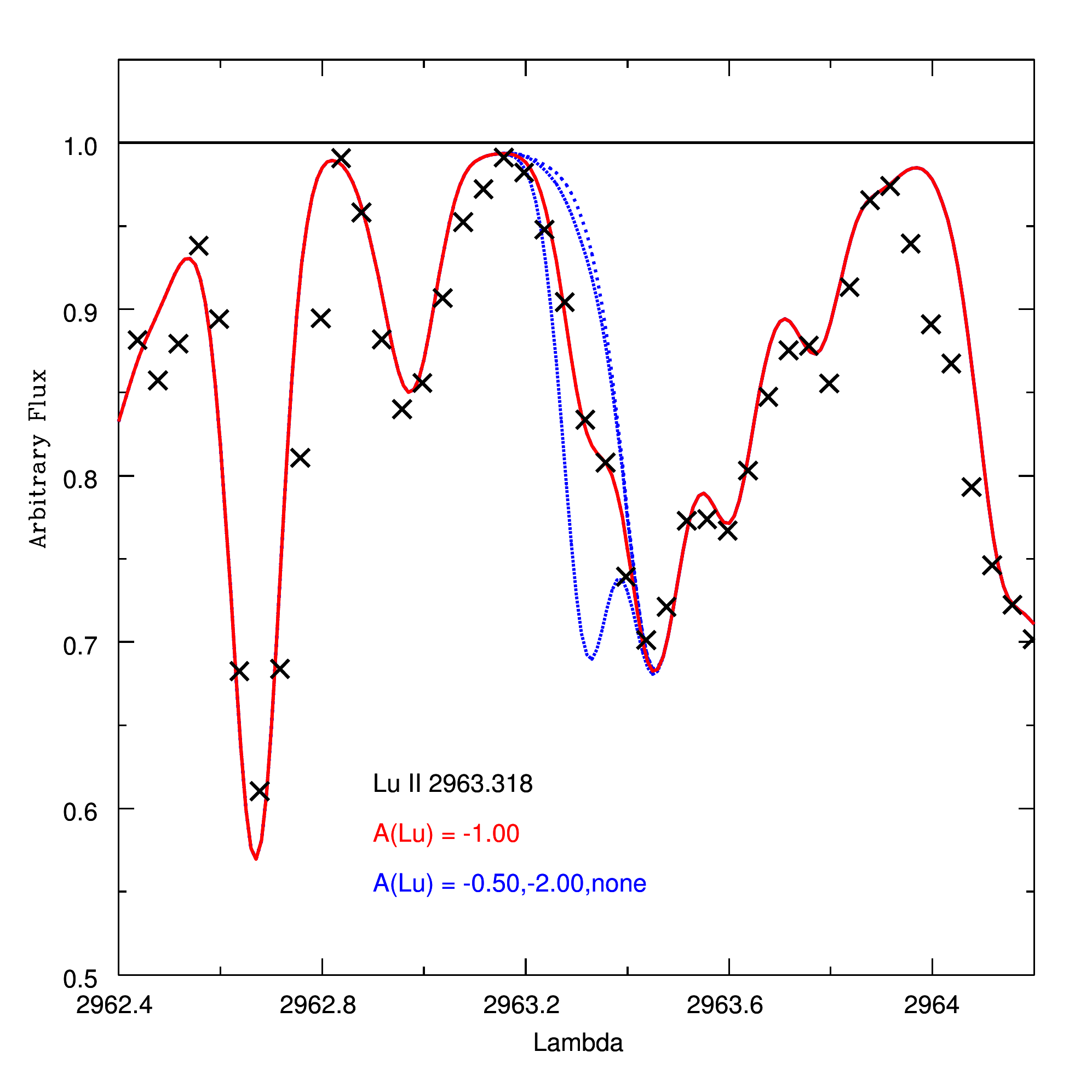}}
\caption{Fit of the observed Lu II 2963.318 {\AA} line in CS 31082-001. Symbols as in Fig. 1.}
\label{Lu2}
\end{figure}

\begin{figure}
%\resizebox{\hsize}{!}{\includegraphics[angle=0]{Hf.pdf}}
\centering
\resizebox{70mm}{!}{\includegraphics{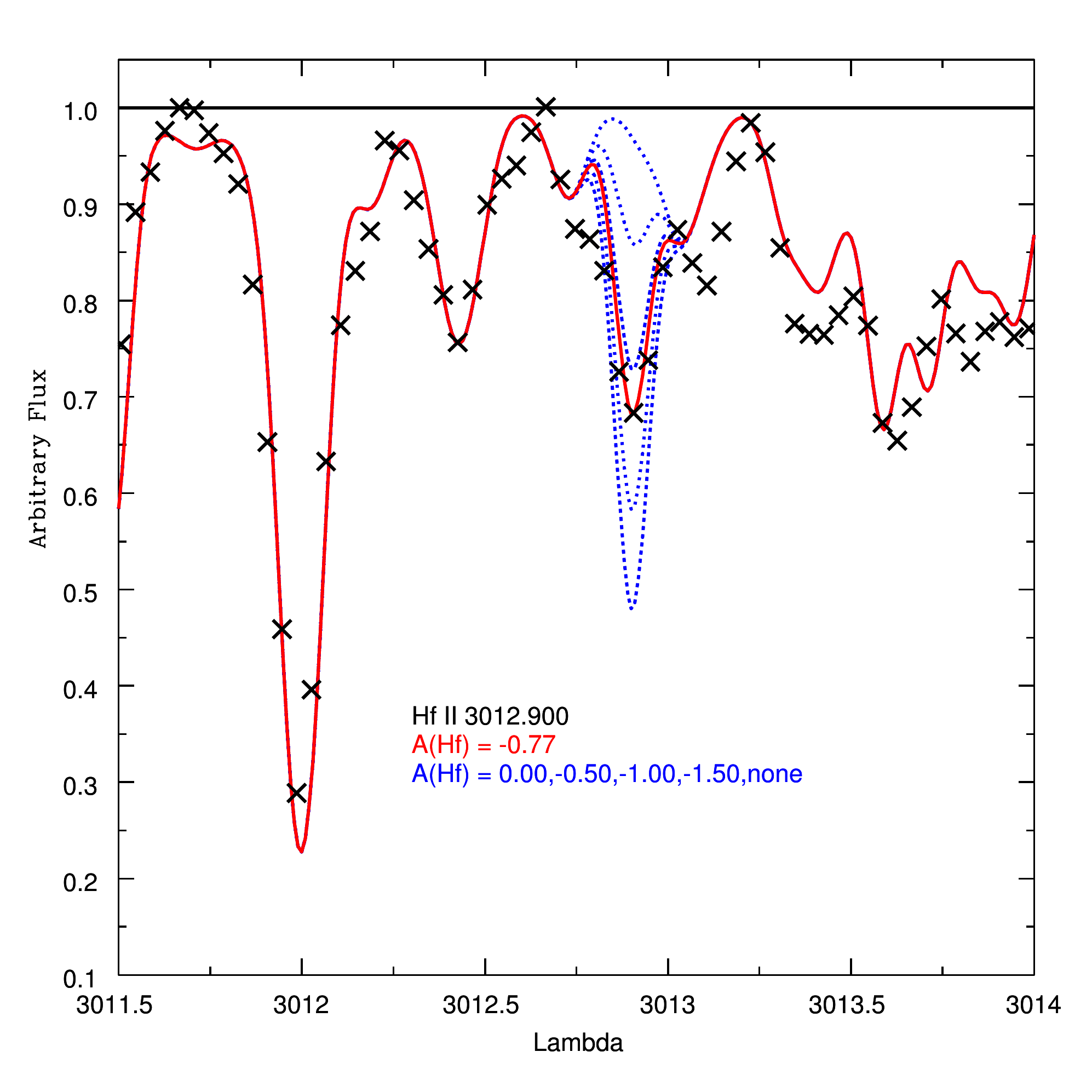}}
\caption{Fit of the observed Hf II 3012.900 {\AA} line in CS 31082-001. Symbols as in Fig. 1.}
\label{Hf}
\end{figure}

\begin{figure}
%\resizebox{\hsize}{!}{\includegraphics[angle=0]{Ta.pdf}}
\centering
\resizebox{70mm}{!}{\includegraphics{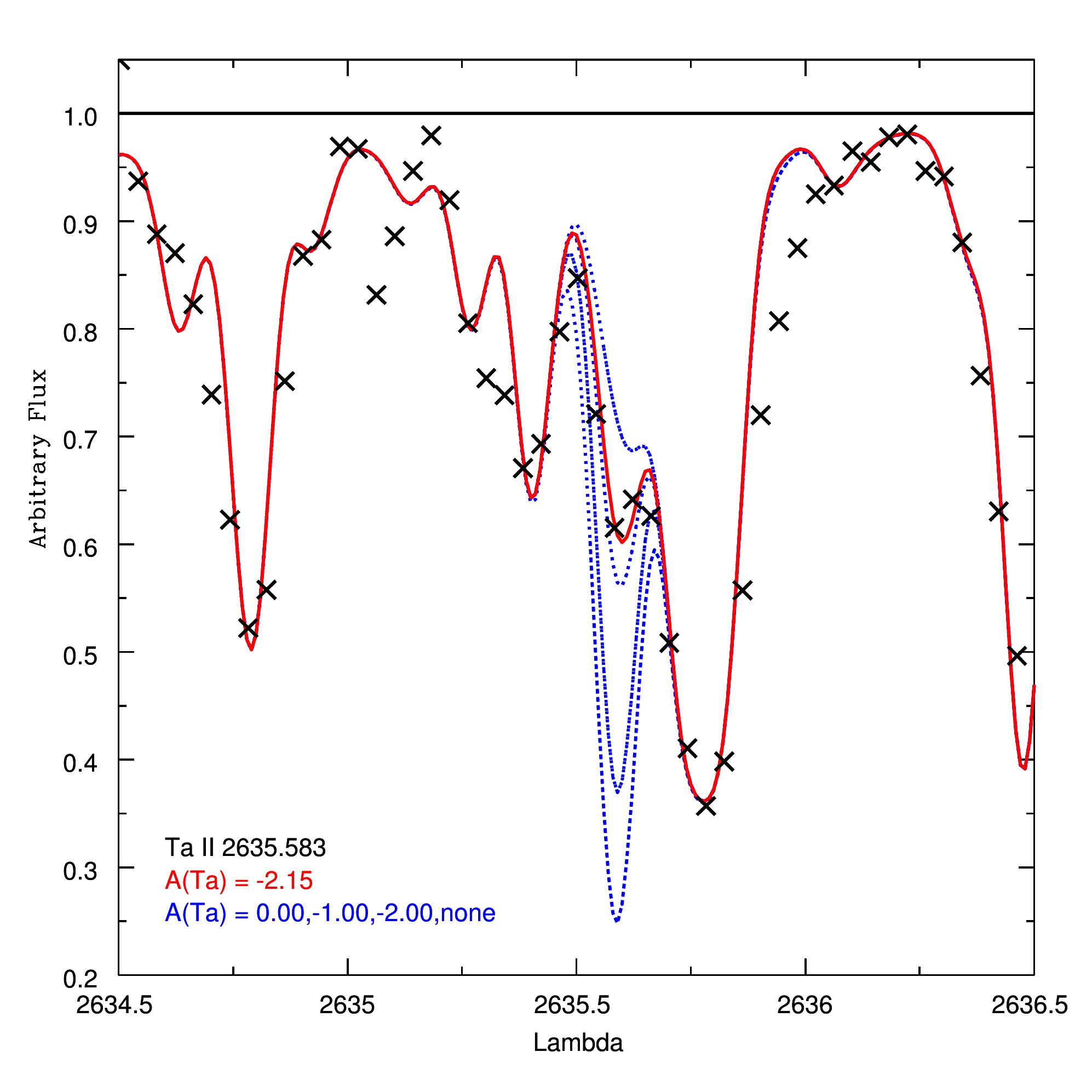}}
\caption{Fit of the observed Ta II 2635.583 {\AA} line in CS 31082-001. Symbols as in Fig. 1.}
\label{Ta}
\end{figure}

\begin{figure}
%\resizebox{\hsize}{!}{\includegraphics[angle=0]{W.pdf}}
\centering
\resizebox{70mm}{!}{\includegraphics{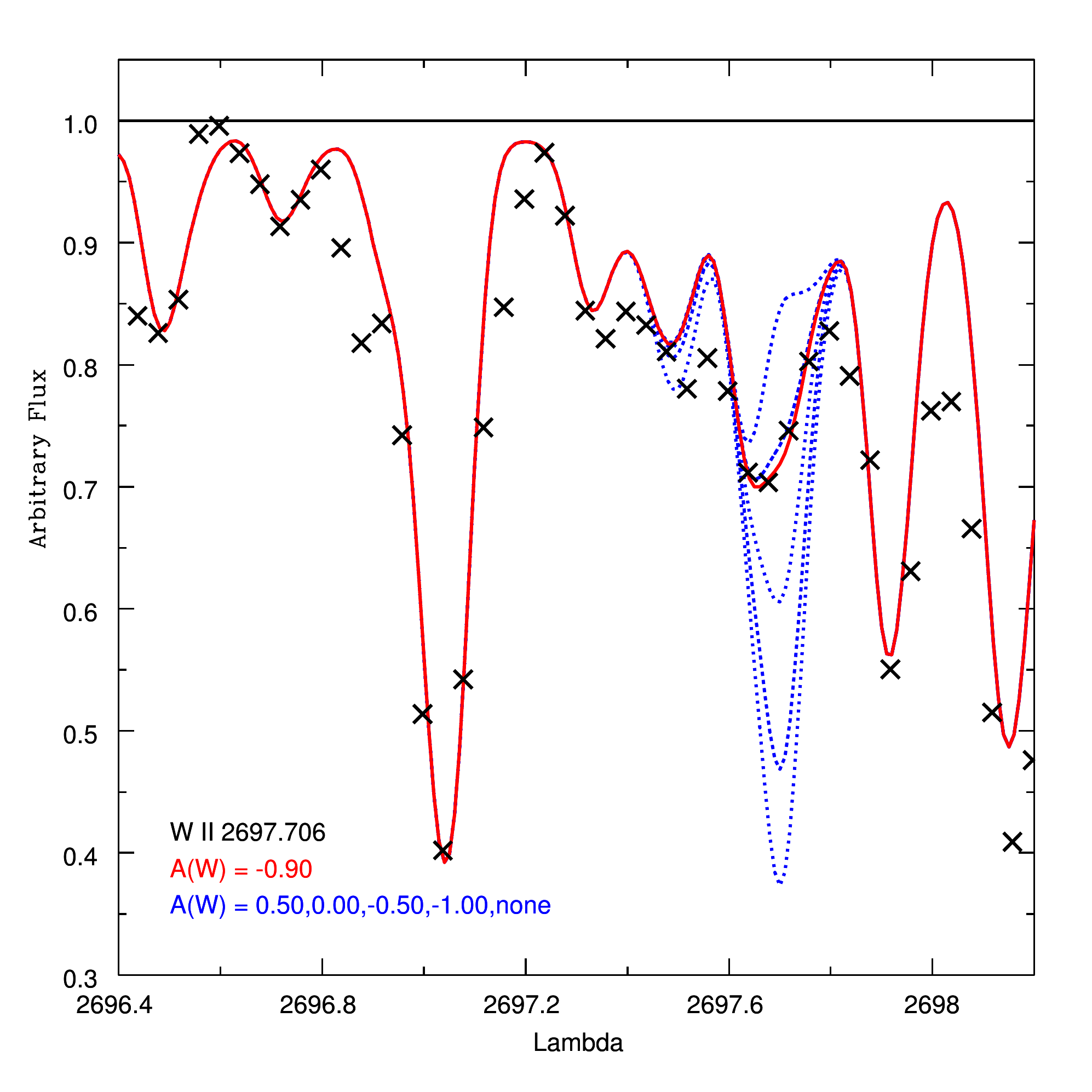}}
\caption{Fit of the observed W II 2697.706 {\AA} line in CS 31082-001. Symbols as in Fig. 1.}
\label{W}
\end{figure}

\begin{figure}
%\resizebox{\hsize}{!}{\includegraphics[angle=0]{Re.pdf}}
\centering
\resizebox{70mm}{!}{\includegraphics{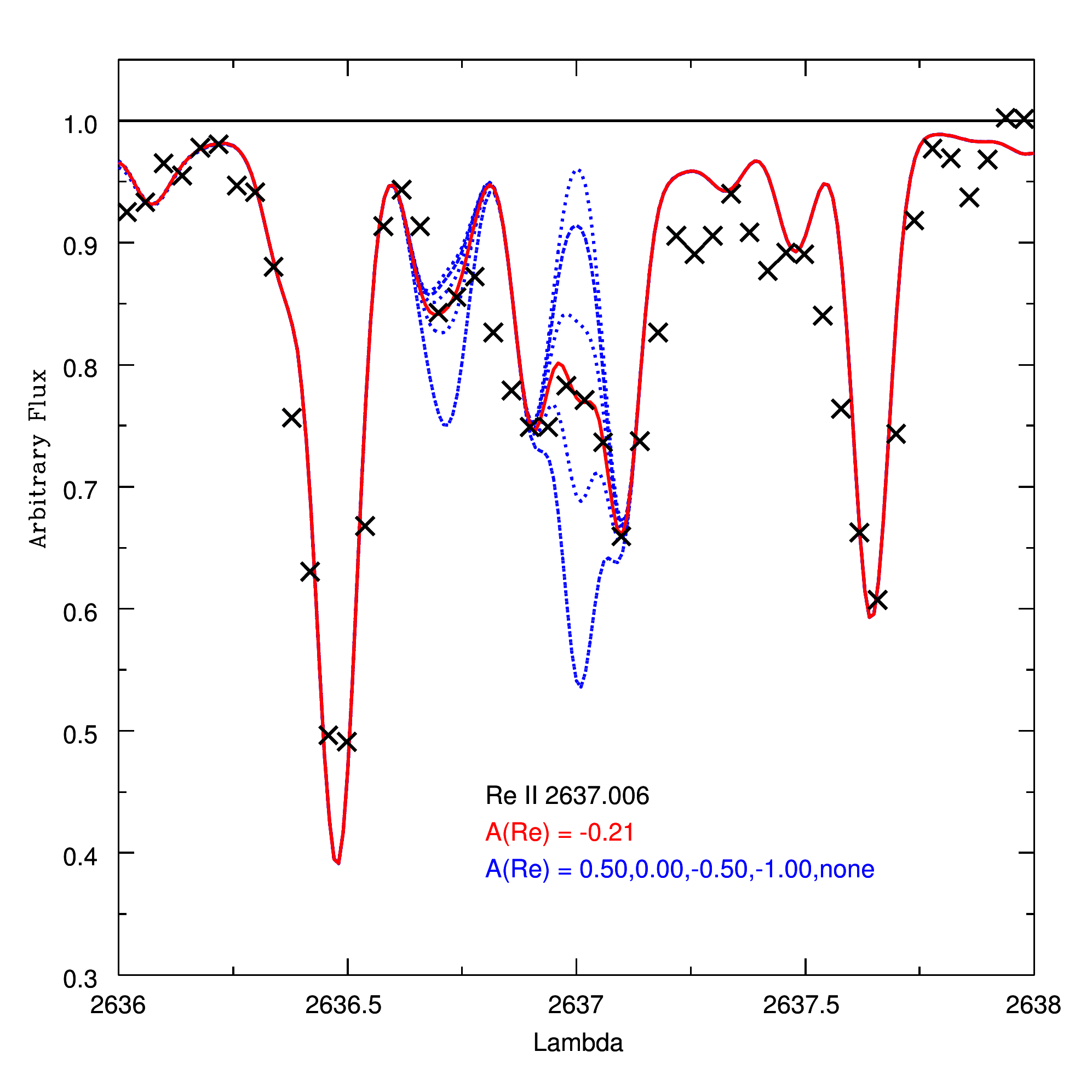}}
\caption{Fit of the observed Re II 2637.006 {\AA} line in CS 31082-001. Symbols as in Fig. 1.}
\label{Re}
\end{figure}

\subsubsection{Ruthenium (Z=44)}

The VALD atomic data for many transitions of Ru I give equivalent widths that are too strong,
 as discussed by \citet{Barbuy2011}, and new 
laboratory oscillator strengths have not been presented in the recent literature. 
However, we tried to use 
one Ru I line to determine a new UV abundance from STIS, and we found A(Ru)=0.65 dex, higher than the previous result 
A(Ru)=0.36$\pm$0.17 dex from \citet{Hill2002}, confirming the impossibility of using the region of the STIS spectra 
for this element.

We calculated the abundance again using the three lines observed in our new UVES spectrum,
and also used by \citet{Hill2002} and we found 
A(Ru)=0.36$\pm$0.12 dex ([Ru/Fe]=1.45), in very good agreement with the previous work.

\subsubsection{Rhodium (Z=45)}

Using three good Rh I lines, we were able to determine A(Rh)=-0.42$\pm$0.12 dex ([Rh/Fe]=1.39) for the abundance of this element. 
Our new result is in complete agreement with A(Rh)=-0.42$\pm$0.16 dex from \citet{Hill2002} and 
this value was adopted as the final abundance.

\subsubsection{Palladium (Z=46)}

Adding the line Pd I 3516.944 {\AA} to the original set of transitions from \citet{Hill2002}, the new UVES spectra give 
A(Pd)=-0.09$\pm$0.07 dex, in good agreement with A(Pd)=-0.05$\pm$0.18 dex from the previous work. Here we have adopted 
our new result [Pd/Fe]=1.18 as the final abundance.

\subsubsection{Silver (Z=47)}

For this element, only the Ag I lines 3280.679 {\AA} and 3382.889 {\AA} were useful with the UVES spectrum, the same ones 
as used by \citet{Hill2002}. 
Our new analysis gives A(Ag)=-0.84$\pm$0.21 dex ([Ag/Fe]=1.15) as the final abundance, in agreement with the previous 
result A(Ag)=-0.81$\pm$0.24 dex. The lower error found in the present work is probably due to the better quality of the 
new UVES spectra in the bluer region.

\subsection{Elements of the second peak}

\subsubsection{Barium (Z=56)}

This is another element studied by \citet{Hansen2012} in CS 31082-001 with a big difference when compared with previous results. They found [Ba/Fe]=1.43, 
which is 0.27 higher than our adopted value from \citet{Hill2002}. While the abundance in this last work was derived from six Ba II lines, 
\citet{Hansen2012} used only two of them, Ba II 4554.03 {\AA} and Ba II 5853.67 {\AA}, leading us to check these results again.

In addition to the comments made in the case of yttrium, we also adopted the same reference as for the hyperfine structure
of barium \citep{McWilliam1998}. 
The final abundance was calculated with the weighting method suggested by the authors to allow us a genuine comparison \citep[for details, see][]{Hansen2012}. 
We found [Ba/Fe]=1.04 with our original atmospheric parameters, while the set from \citet{Hansen2012} gave the relative abundance [Ba/Fe]=1.30. 
The difference between the results is 0.26, very close to the previous difference. Since the barium lines are strong, we explain this difference as effect of the 
microturbulence velocity, and we decided to keep the result from \citet{Hill2002} as the final abundance. 
Considerations about the NLTE corrections on this result can be found in section 3.5.

\subsubsection{Cerium (Z=58)}

By using ten new lines of Ce II we were able to determine the abundance A(Ce)=-0.31$\pm$0.10 dex, in very good agreement with 
the value from \citet{Hill2002}. However, \citet{Sneden2009} determined a more precise abundance using 38 transitions 
for this element, leading us to adopt their result A(Ce)=-0.29$\pm$0.05 dex ([Ce/Fe]=1.03) as the final abundance.

\subsubsection{Neodymium (Z=60)}

We derived an abundance A(Nd)=-0.21$\pm$0.10 dex from the 18 useful Nd II lines in the region covered by the new UVES spectra, 
in agreement with A(Nd)=-0.13$\pm$0.13 dex found by \citet{Hill2002} and with A(Nd)=-0.15$\pm$0.05 dex by \citet{Sneden2009}, 
despite the lower absolute value. In fact, even using only the subset of lines with improved atomic data we found 
A(Nd)=-0.24$\pm$0.10 dex, quite similar to the complete set. Other authors use sets of lines covering the entire optical region, 
and we adopted the result from \citet{Sneden2009} [Nd/Fe]=1.33 as the final abundance, since they have used the most complete 
line list for this element.

\subsubsection{Samarium (Z=62)}

This element presents 23 useful lines in the region studied, and we found the abundance A(Sm)=-0.42$\pm$0.09 dex from the 
set of data, in complete agreement with A(Sm)=-0.42$\pm$0.05 dex ([Sm/Fe]=1.51) from \citet{Sneden2009}. We decided to keep this 
last result as the adopted abundance since the number of lines used is considerably higher, making the error smaller.

\subsubsection{Europium (Z=63)}

After checking more than 20 profiles, we derived the abundance A(Eu)=-0.75$\pm$0.22 dex from the best Eu II 2906.669 {\AA}, in 
agreement with A(Eu)=-0.72$\pm$0.05 dex from \citet{Sneden2009}. Fig. \ref{Eu} shows our fit to this line. 
We finally adopted the value from \citet{Sneden2009} as the more reliable result ([Eu/Fe]=1.69) given the higher number of lines used.

\subsubsection{Gadolinium (Z=64)}

This element shows several available lines in this region, and we derived its abundance using 32 Gd II lines. 
We found A(Gd)=-0.29$\pm$0.09 dex, in agreement with A(Gd)=-0.21$\pm$0.05 
from \citet{Sneden2009}. 

In this work we are using the same references as \citet{Sneden2009} for new atomic data, but not all the transitions 
have been updated by \citet{Den Hartog2006}, so we adopted the value from \citet{Sneden2009} as more reliable ([Gd/Fe]=1.61).

\subsubsection{Terbium (Z=65)}

This is the most problematic element with respect to compatibility between previous abundance results. \citet{Hill2002} found 
A(Tb)=-1.26$\pm$0.12 dex with $\sigma$=0.07 from seven optical lines using the UVES spectra, while \citet{Sneden2009} found 
A(Tb)=-1.01$\pm$0.05 dex with $\sigma$=0.04 from nine lines using updated oscillator strengths from \citet{Lawler2001c}. 
In this work we were able to use three Tb II lines to derive the abundance A(Tb)=-1.00$\pm$0.14 dex. For the line 2934.802 {\AA} 
no new gf-value is available so it was excluded from the average, despite a good fit.

Our new result confirms the lower abundance found by \citet{Sneden2009}, and we decided to keep [Tb/Fe]=1.64 as the more 
reliable abundance since the authors used a bigger set of lines.

\subsubsection{Dysprosium (Z=66)}

From 26 profiles in the region studied, we determined the abundance A(Dy)=-0.16$\pm$0.09 dex. By selecting only those lines with new atomic data, 
the abundance found is A(Dy)=-0.12$\pm$0.09 dex, in agreement with A(Dy)=-0.07$\pm$0.05 dex from \citet{Sneden2009}.
 We finally readopted the value 
resulting from the largest set of lines [Dy/Fe]=1.73 from \citet{Sneden2009} as the final abundance.

\subsubsection{Erbium (Z=68)}

After checking several Er lines, we were able to derive a new abundance A(Er)=-0.31$\pm$0.09 dex from 17 good lines 
of Er II. Fig. \ref{Er} shows an example of fit to an Er II line. 
Our result agrees with A(Er)=-0.27$\pm$0.15 dex found by \citet{Hill2002} and with A(Er)=-0.30$\pm$0.05 dex found by 
\citet{Sneden2009}, which used the same reference for atomic data. However, the last authors derived their result from 19 
erbium lines, with $\sigma$=0.04, which combined with our higher uncertainty, leads us to keep the value [Er/Fe]=1.67 found by 
\citet{Sneden2009} as the most reliable.

\subsubsection{Thulium (Z=69)}

From several Tm lines, we were able to use the nine best lines to derive an abundance A(Tm)=-1.18$\pm$0.10 dex. Our result
agrees with A(Tm)=-1.24$\pm$0.13 dex by \citet{Hill2002} and A(Tm)=-1.15$\pm$0.05 dex by \citet{Sneden2009}. 

\subsubsection{Lutetium (Z=71)}

Another new element derived in CS 31082-001, the adopted abundance is A(Lu)=-1.08$\pm$0.13 dex ([Lu/Fe]=1.73), 
obtained from the mean of the results derived from the three best Lu II lines (see Figs. \ref{Lu1} and \ref{Lu2}) 
from our set of data. 

\subsubsection{Hafnium (Z=72)}

After checking more than 60 lines in the region studied, we derived a hafnium abundance A(Hf)=-0.73$\pm$0.11 dex from five lines 
that were not affected by unidentified lines or other problems. Figure \ref{Hf} shows an example of fit to an Hf II line. Our result is in good agreement 
with A(Hf)=-0.72$\pm$0.05 dex ([Hf/Fe]=1.33) from \citet{Sneden2009}, which used new atomic data from \citet{Lawler2007}. 
We decided to keep the last result as the most reliable abundance.

\subsubsection{Tantalum (Z=73)}

As discussed in \citet{Barbuy2011}, this is another element that presents lines that are too strong for many transitions 
from the VALD database, and also an element that has not been analyzed in this star. We found new oscillator strengths for 
five Ta I lines in \citet{Fivet2006}, but all of them were discarded for typical problems of UV region 
(as the definition of the continuum). 
However, we tried to use three other Ta II lines present in the spectrum.

Ta II 2685.190 {\AA} gives A(Ta)=-2.80 dex, but the synthetic profile of the line is really strong and probably its 
oscillator strength value is not correct. On the other hand, Ta II  2832.702 {\AA} gives us an abundance A(Ta)=-1.05 dex with a 
good fit, despite the line being weak. Another abundance indicator is the Ta II 2635.583 {\AA} line, shown in Fig. \ref{Ta}, 
which yields A(Ta)=-2.15 dex. The final tantalum abundance is the average of the last two lines, A(Ta)=-1.60$\pm$0.23 dex 
([Ta/Fe]=1.47).

\subsubsection{Tungsten (Z=74)}

From our set of W II lines, most of them have new atomic data presented in \citet{Nilsson2008}. We derived the abundance 
A(W)=-0.90$\pm$0.24 dex ([W/Fe]=0.92) using our best line W II 2697.706 {\AA}, shown in Fig. \ref{W}, also another first determination 
in this star.

\subsubsection{Rhenium (Z=75)}

For the heaviest element analyzed in this work and also a new one for this star, the abundance A(Re)=-0.21$\pm$0.21 dex 
([Re/Fe]=2.45) was derived from our two best lines Re I 2930.613 {\AA} and Re II 2637.006 {\AA} (see Fig. \ref{Re}).
Together with tungsten, rhenium abundance is extremely important for studying the transition region between the second and the third 
peaks of the r-process.

\subsection{NLTE and tridimensional corrections}

\citet{Andrievsky2009,Andrievsky2011} reanalyzed the sample of the EMP stars previously studied in the framework 
of the ESO Large program ``First Stars'', including CS 31082-001, determining in particular the abundance of Sr and Ba 
based on NLTE computations. For Sr the NLTE abundance is 0.2 dex lower than the LTE value found in 
\citet{Hill2002} ([Sr/Fe]$_{NLTE}$=0.53), while a stronger correction to the Ba abundance was found, and the NLTE value is A(Ba)=0.00 dex ([Ba/Fe]$_{NLTE}$=0.76).

Recently, \citet{Mashonkina2012} have considered
 the ultraviolet overionization to calculate the NLTE abundance of Pb in cool stars for the Pb I line 4057 {\AA}. 
In the case of CS 31082-001 the corrected value is A(Pb)=+0.01 ([Pb/Fe]$_{NLTE}$=0.94), substantially higher than the previous LTE abundances A(Pb)=-0.55 from 
\citet{Plez2004} and A(Pb)=-0.65 from \citet{Barbuy2011}. For completeness, we also present the NLTE abundance for europium calculated 
by \citet{Mashonkina2012}, 0.06 dex higher ([Eu/Fe]$_{NLTE}$=1.75) than the best LTE value A(Eu)=-0.72 dex from \citet{Sneden2009}.

For the other heavy elements NLTE corrections are not available. It would be particularly interesting to check the NLTE effects on the Ge abundances, 
since the main abundance indicator for this element is a transition from the non ionized state. 

It is well known that NLTE calculations relative to LTE have effects on the abundances that are counterbalanced by taking
3D modeling into account, therefore both effects should be computed at the same time. 
In fact, since the lead abundance is an important calibration point for zero-age r-process abundance distribution models 
\citep{Schatz2002,Wanajo2007}, and the 
NLTE correction found in the literature is high enough to have implications for the discussion of
 r-process models for the heaviest n-capture elements in this star, 
we calculated the 3D correction for the abundance of this element.

For computing of the 3D correction we used a hydrodynamical model computed with
 the code CO5BOLD \citep{Freytag2002,Freytag2012} with parameters 5020/2.5/-3/0. 
The model has a resolution of $160\times 160\times 200$, and its dimension is $573215\times 573214\times 245362~ {\rm km}^3$. 
Twenty representative snapshots have been selected, 
covering 156 h in time. For the opacity, based on the MARCS stellar atmosphere package \citep{Gustafsson2008},
 the absorption coeffcients were averaged in six bins. 
The plane parallel 1D$_{\rm LHD}$ model was used as reference model, computed with the LHD code that shares the micro-physics and opacity with the CO5BOLD code.

The 3D correction is defined as A(3D)-A(1D$_{\rm LHD}$) \citep[for details, see][]{Caffau2011}. The line formation computations were done with 
Linfor3D\footnote{http://www.aip.de/$\sim$mst/Linfor3D/linfor\_3D\_manual.pdf}. Compared with the 1D LTE value, the 3D effect presents a correction of $\Delta$A(Pb)=-0.21 dex, 
which together with the NLTE correction, gives a new lead abundance of A(Pb)=-0.20 dex ([Pb/Fe]$_{NLTE}$=0.73). 
As discussed by \citet{Spite2012}, weaker lines form in deeper layers, where the 
granulation velocities and intensity contrast are higher, which is a possible explanation for this significant correction in the lead abundance.

It is important to note that our correction is not a full 3D NLTE computation, 
which has only been performed for Li I and O I so far. But while these complete models seem to be important in the Li case 
\citep{Asplund2003,Barklem2003,Cayrel2007,Sbordone2010}, the NLTE corrections for O I are quite similar in the 3D and 1D cases, 
at least with the solar parameters \citep{Asplund2004}.

\section{Discussion}

\begin{figure}
%\resizebox{\hsize}{!}{\includegraphics[angle=0]{figura3.pdf}}
\centering
\resizebox{95mm}{!}{\includegraphics{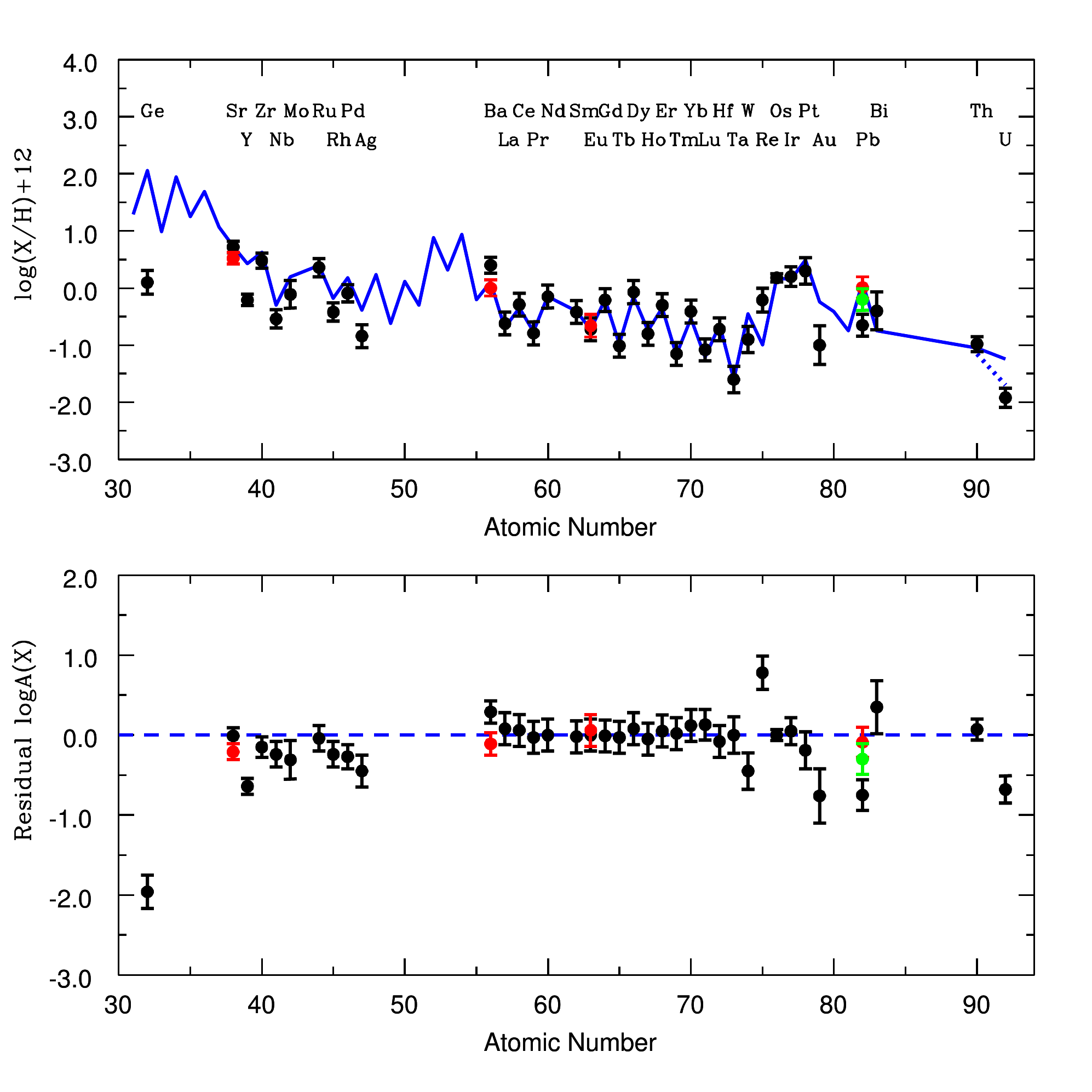}}
\caption{Solar r-process abundance values. Top: deconvolution from \citet{Simmerer2004} scaled to Eu (solid line) 
compared with the new complete observed abundances in CS 31082-001 (black dots and respective error bars). Radioactive element (Th and U) abundances 
are corrected for radioactive decay since the formation of the solar system. The dotted line shows the abundances observed 
today for these two species. Bottom: Abundance residuals. 
NLTE abundances for some elements (red dots and respective error bars) are compared with the LTE results. For Pb, the green symbol represents the NLTE+3D corrected value.}
\label{sim}
\end{figure}

\subsection{Comparison with the solar system r-process pattern}

The abundances in CS 31082-001 as determined from previous works \citep{Hill2002,Sneden2009} and 
from the present analysis are compared with each other in Table \ref{abs}. 
A comparison of the observed abundance pattern with the scaled solar system r-process abundances using the deconvolution from 
\citet{Simmerer2004} is shown in Fig. \ref{sim}, together with the residual values. 
The figure includes the results from \citet{Barbuy2011} for the third-peak r-process elements.
For the radioactive elements U and Th, the abundance corrected for radioactive decay since the formation of the 
solar system, and the abundances observed today are shown. 
In addition, the available NLTE abundances for some elements (red dots and respective error bars) are compared with the LTE results, 
and in the case of lead we also present the new NLTE+3D corrected abundance as the green symbol.

Fig. \ref{sim} shows the NLTE corrected abundances compared with the LTE result and with the solar system r-process abundances, 
from which it is possible to see that the NLTE corrections from \citet{Andrievsky2011} make the Sr abundance more similar to the trends of the other elements from the first peak. 
One can also see from Fig. \ref{sim} that the NLTE Ba abundance from \citet{Andrievsky2009} is in good agreement with the Solar System r-process pattern. 
Finally, the comparison shows that the new NLTE+3D lead abundance is in better agreement but still lower than the solar system r-process value.

As discussed in some detail in the last section, there is clear compatibility of the observational results for each element. 
Furthermore, Fig. \ref{sim} shows that the n-capture element distribution in the star matches a scaled solar r-process pattern 
very well, from Ba (Z=56) through the third r-process peak. 
This well-known result in the context of metal-poor stars enriched in r-process 
elements had led some authors to argue that this extremely close agreement is evidence of the robust nature of the r-process, 
operating in much the same manner over the lifetime of the Galaxy. Our new abundances for lutetium and tantalum follow this trend, 
but while the tungsten value seems to be under-solar, the rhenium is overabundant. The disagreement can originate in a break down of the 
universality for Z = 74-75, from our spectrum and/or the atomic data, but another possible error source is the solar system r-process deconvolution \citep{Goriely1999}.

\citet{Sneden2008} present  an abundance comparison in their Fig. 11 with extensive elemental data for six r-process-rich stars, 
including CS 31082-001, showing exactly the consistency between the abundances of the heavier stable n-capture elements and the 
solar system r-process abundance distribution. This group of stars has been identified as standard templates to characterize the 
r-process nucleosynthesis pattern. At the same time, the comparison shows that the match between the stellar r-process abundances 
and the scaled solar system r-process pattern does not extend to the lighter heavy elements, and Fig. \ref{sim} shows this 
well known result for the specific case of CS 31082-001 with the new abundance determination of molybdenum and germanium. 
Recently, \citet{Roederer2010a} and \citet{Cowan2011} have proposed that, similarly to the need of having several mechanisms
operating, in order 
to explain the origin of the lightest trans-Fe elements, a simple linear combination of the scaled solar system s-process 
and r-process is an inadequate description of some of the heavy n-capture elements, when a precise deconvolution is desired. 
On the other hand, the standard method of computing r-process residuals by solar deconvolution is still adequate for assessing the relative dominance of the s- or 
r-process in a general sense. 

In addition, as discussed in \citet{Barbuy2011}, we also see disagreements between observation and theory among the heaviest 
elements from the third r-process peak and actinides; notably, the high ages derived from the U/Os, U/Ir, and U/Pt ratios in the radioactive 
chronometry context would indicate that the nuclear data and/or astrophysical modeling of elements production are in need of improvement. On the other hand, 
 that a strong actinide boost is observed in CS 31082-001, but not in other r-II stars like HE 1523-0901, suggests 
that the production of the heaviest elements in the r-process site(s) may be more complex than supposed so far.

\begin{figure}
%\resizebox{\hsize}{!}{\includegraphics[angle=0]{wanajo.pdf}}
\centering
\resizebox{95mm}{!}{\includegraphics{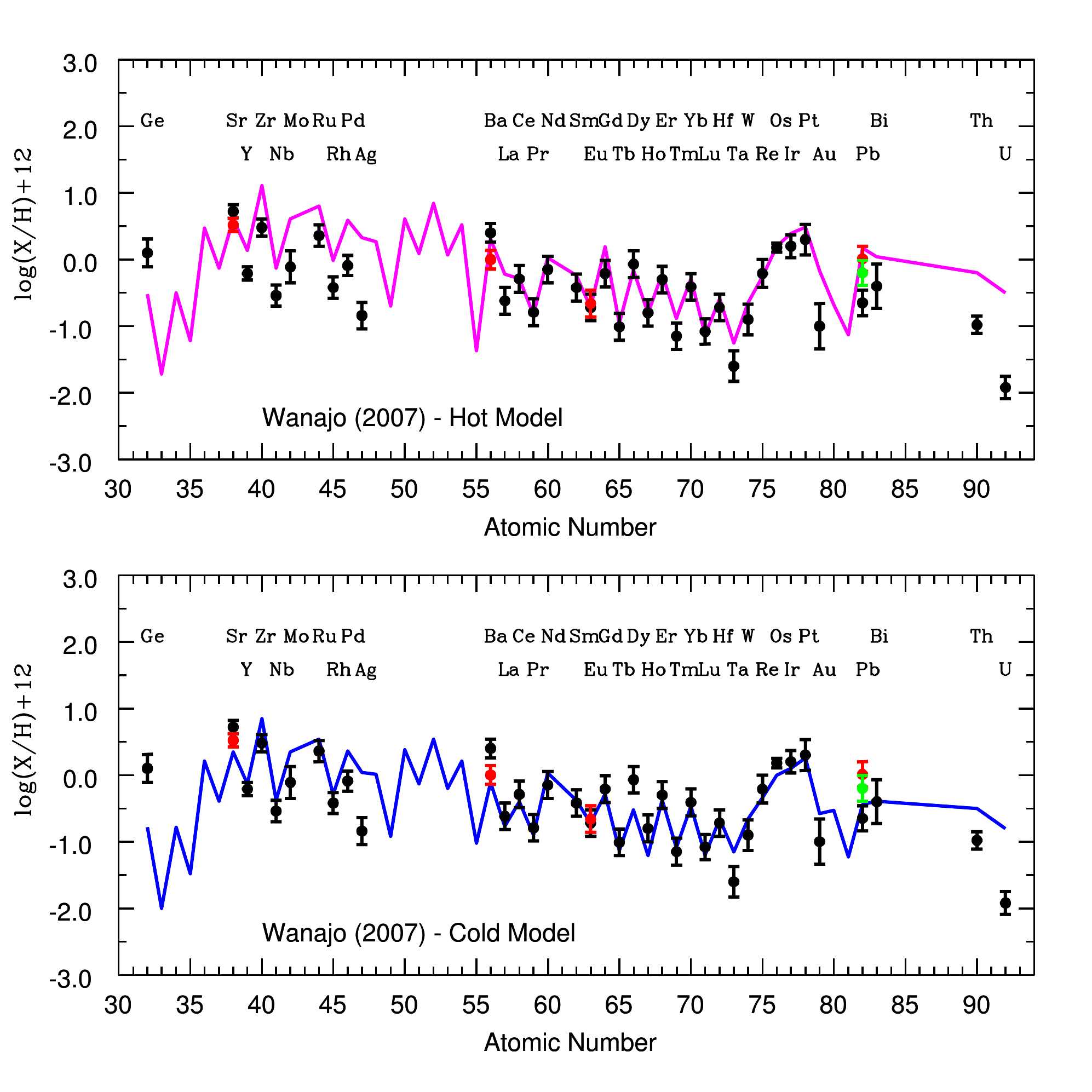}}
\caption{Predicted abundance patterns for the hot (upper) and cold (lower) models by \citet{Wanajo2007} (solid lines), 
compared with the new complete observed abundances in CS 31082-001. Symbols as in Fig. 15.}
\label{wanajo}
\end{figure}

\begin{figure}
%\resizebox{\hsize}{!}{\includegraphics[angle=0]{erro.pdf}}
\centering
\resizebox{95mm}{!}{\includegraphics{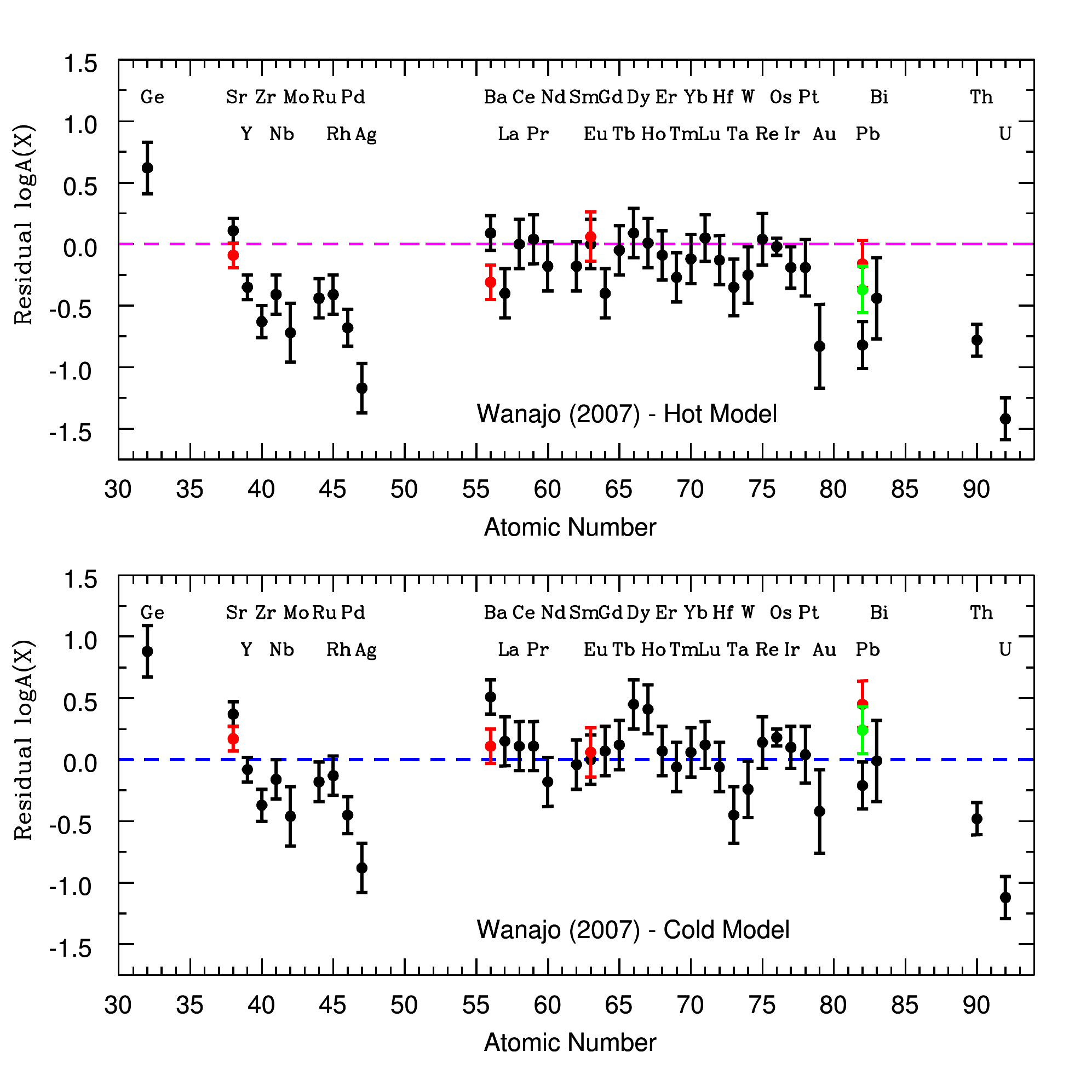}}
\caption{Abundance residuals of elements in CS 31082-001 from the two \citet{Wanajo2007} model predictions. Symbols as in Fig. 15.}
\label{wanajo_res}
\end{figure}

\subsection{Comparison with r-process models}

Figure \ref{wanajo} compares the predictions of the hot and cold models by \citet{Wanajo2007} with the observed 
abundances in CS~31082-001, including the new elements derived in this work, while Fig. \ref{wanajo_res} shows the corresponding 
residuals. The abundances obtained from these models of nucleosynthesis calculations are based on supernovae neutrino 
wind models with updated nuclear input data \citep[based on the HFB-9 model of][]{Goriely2005}. 
These data update the older calculations by \citet{Wanajo2002}, who adopted a cold r-process that proceeds with competing (n, $\gamma$) and $\beta$-decays, but without 
($\gamma$, n) decays when the temperature drops down to 1.0 $\times$ 10$^{8}$ K. 
This differs from the traditional (hot) r-process at a temperature of 1.0 $\times$ 10$^{9}$ K, where the (n, $\gamma$)-($\gamma$, n) 
equilibrium remains a good approximation during r-processing. 

The hot model fits many of the second-peak elements well, but fails for the first-peak elements and the 
heaviest third-peak elements. The cold model gives an overall better fit, except for Ba, Dy, Ho, and Os. 
This might suggest that the abundance pattern results from nucleosynthetic processes in 
several different physical conditions. It is also worth noting that both the hot and cold 
models fit  our new Lu, W, and Re 
abundances very well (for two of which poor agreement with the solar
 r-pattern can be seen in Fig. \ref{sim}), despite the failure to fit the Ta abundance.
We note that it is the first time that abundances of elements from the region 
between the second and third peaks are given for an EMP star. 
An important difference between the models is the behavior of the abundance pattern for the heaviest elements of the third peak. 
While the traditional ``hot model'' from \citet{Wanajo2002} produces abundances of gold, lead, and bismuth
that are substantially higher than the observed values, 
the cold model seems to be better for describing this region, 
as discussed in \citet{Barbuy2011}. Albeit more subtly, the new NLTE+3D Pb abundance leads us 
to the same conclusion. It would be interesting to check these corrections with the other 
elements in this region, but NLTE corrections for gold and bismuth 
are not available in the literature, any more than the 3D corrections.

Recently, \citet{Farouqi2010} have investigated the termination point of charged-particle freezeout, and they define a maximum 
entropy for a given expansion velocity and electron abundance (Y$_{e}$), beyond which the seed production of heavy elements fails 
owing to the very low matter density. They also investigated an r-process subsequent to the charged-particle freeze-out, analyzing 
the impact of nuclear properties from different theoretical mass models on the final abundances. 
They find it is possible to coproduce the light p-, s-, and r-process isotopes between Zn (Z = 30) and Ru (Z = 44) at 
Y$_{e}$ in the range 0.450-0.498 and low entropies of S$<$100-150 $k_{B}$ (Boltzmann constant) per nucleon 
\citep[see also similar discussion in][]{Hoffman1996,Wanajo2006}. 
They also show that for Y$_{e}$ slightly below 0.50, only the mass region below the mass number A=130 peak can be formed, and the classical “main” r-process region 
up to the full third peak requires somewhat more neutron-rich winds. Figure \ref{faro} shows the results from 
\citet{Farouqi2010} with Y$_{e}$=0.498 and Y$_{e}$=0.482, compared with the observed 
abundances in CS 31082-001. The calculations were performed with a selected constant 
expansion velocity of V$_{exp}$ = 7500 km s$^{-1}$, and for each Y$_{e}$ the superposition 
of the entropies spans from S = 5 $k_{B}$/nucleon to the maximum entropy S$_{final}$(Y$_{e}$)$\sim$300 $k_{B}$/nucleon.
One can see that the whole mass region from Sr up to Th can be fitted by using different parameters, 
in agreement with the need for more than one site for the r-process and/or different conditions into the same enviroment. 
This study can be seen as a generalization of the hot and cold models from \citet{Wanajo2007}, since the parametrization should reach the entire range of possibilities.

\subsection{Origin of germanium}

Another important approach to understanding the origin of the elements is to check the evolution of their abundances as a function of the metallicity, from a sample of star. 
However, for some elements, the number of stars with determined abundances is still reduced by the difficulty of detection.
 That is the case of germanium and molybdenum. 
\citet{Cowan2005} used a sample of ten metal-poor Galactic halo stars with measures of Ge to analyze the behavior of this light 
element compared to the principal r-process patterns, represented by the Eu abundance. The sample includes the r-poor star HD 122563. 
We reproduce the original comparison including CS 31082-001 in the sample with our new Ge abundance. 
The result is shown in the Fig. \ref{cowan1}, and one can see that our star is the most metal-poor object in the sample (Fig. \ref{cowan1} top), and the most enriched with 
r-process elements (Fig. \ref{cowan1} bottom). 
The Ge abundance is correlated with metallicity, but seems to be uncorrelated with the r-process elements. The authors also 
discuss that while n-capture processes are important for Ge production in solar system material, these abundance comparisons 
immediately suggest a different origin for this element early in the history of the Galaxy. 

\begin{figure}
%\resizebox{\hsize}{!}{\includegraphics[angle=0]{faro.pdf}}
\centering
\resizebox{85mm}{!}{\includegraphics{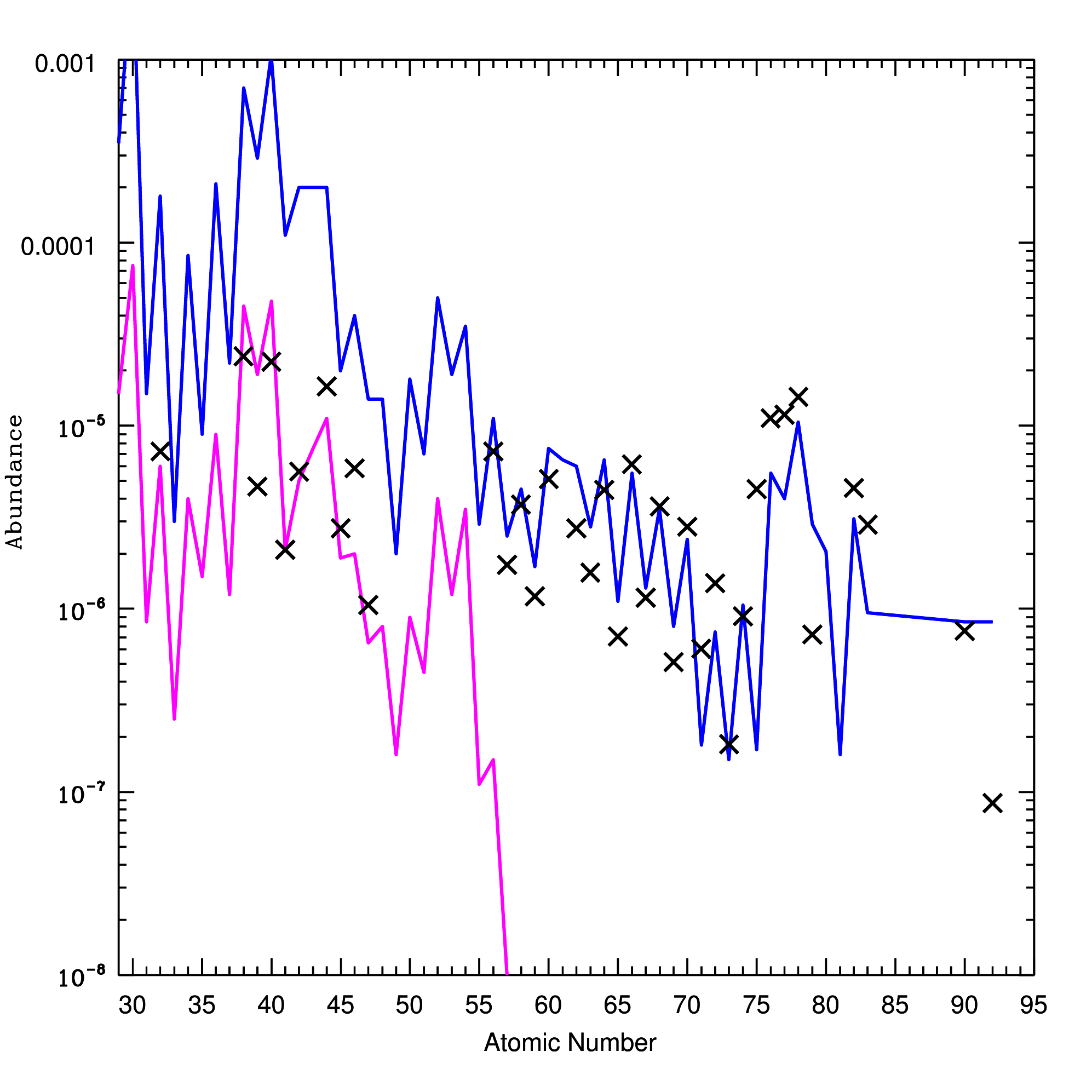}}
\caption{Comparison of the new complete observed abundances in CS 31082-001 (crosses) with yields from \citet{Farouqi2010}, 
using Y$_{e}$ of 0.498 (magenta solid line) and 0.482 (blue solid line). For each Y$_{e}$ the superposition 
of the entropies spans from S = 5 $k_{B}$/nucleon to the maximum entropy S$_{final}$(Y$_{e}$)$\sim$300 $k_{B}$/nucleon.}
\label{faro}
\end{figure}

\begin{figure}
%\resizebox{\hsize}{!}{\includegraphics[angle=0]{faro.pdf}}
\centering
\resizebox{85mm}{!}{\includegraphics{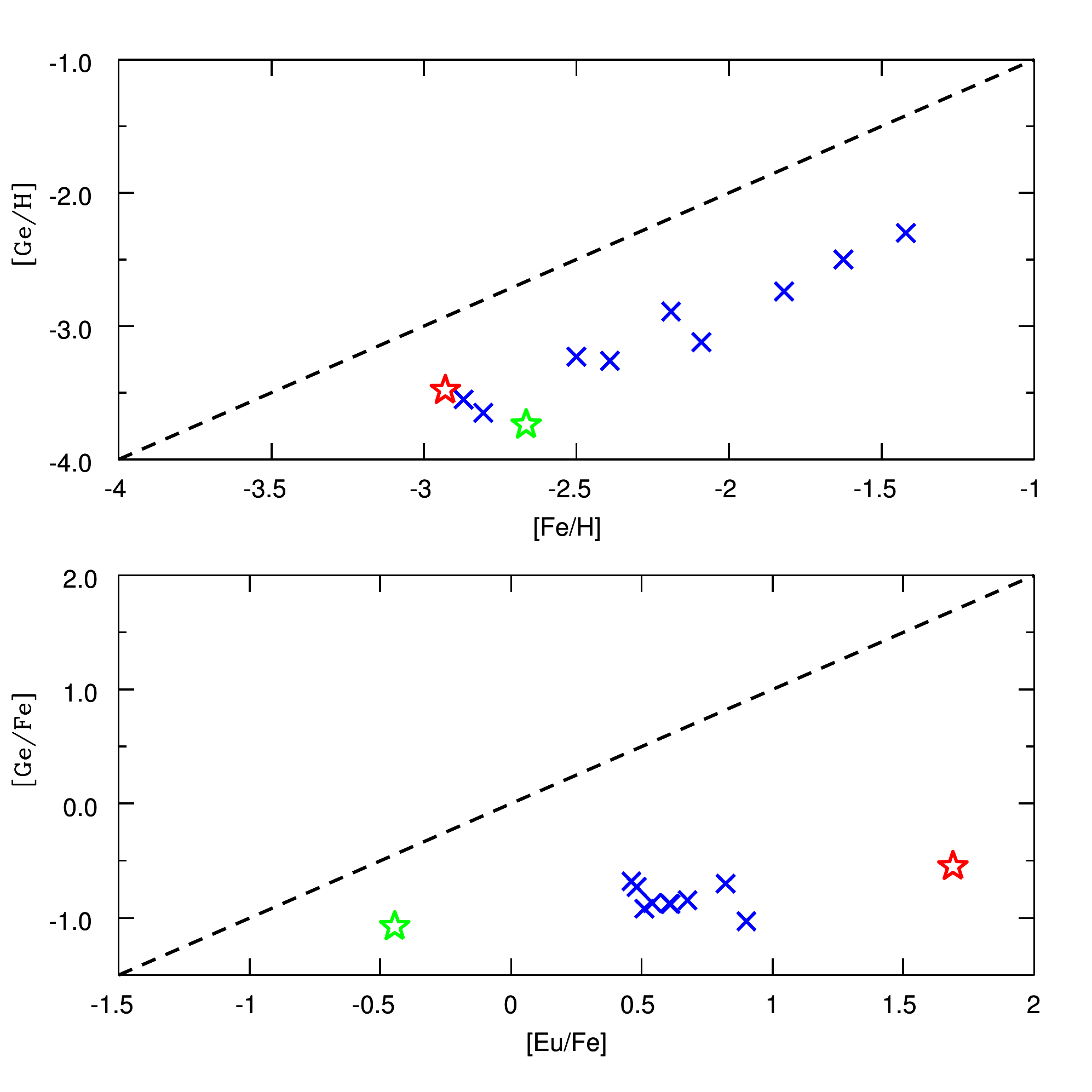}}
\caption{Relative abundances [Ge/H] displayed as a function of [Fe/H] metallicity (top) and correlation between the 
abundance ratios [Ge/Fe] and [Eu/Fe] (bottom). The blue symbols represent the 
original data from \citet{Cowan2005} and our new abundances for CS 31082-001 is marked as the red star. The r-poor HD 122563 is marked as the green star. The dashed line 
indicates the solar abundance ratio of these elements.}
\label{cowan1}
\end{figure}

\begin{figure}
%\resizebox{\hsize}{!}{\includegraphics[angle=0]{faro.pdf}}
\centering
\resizebox{85mm}{!}{\includegraphics{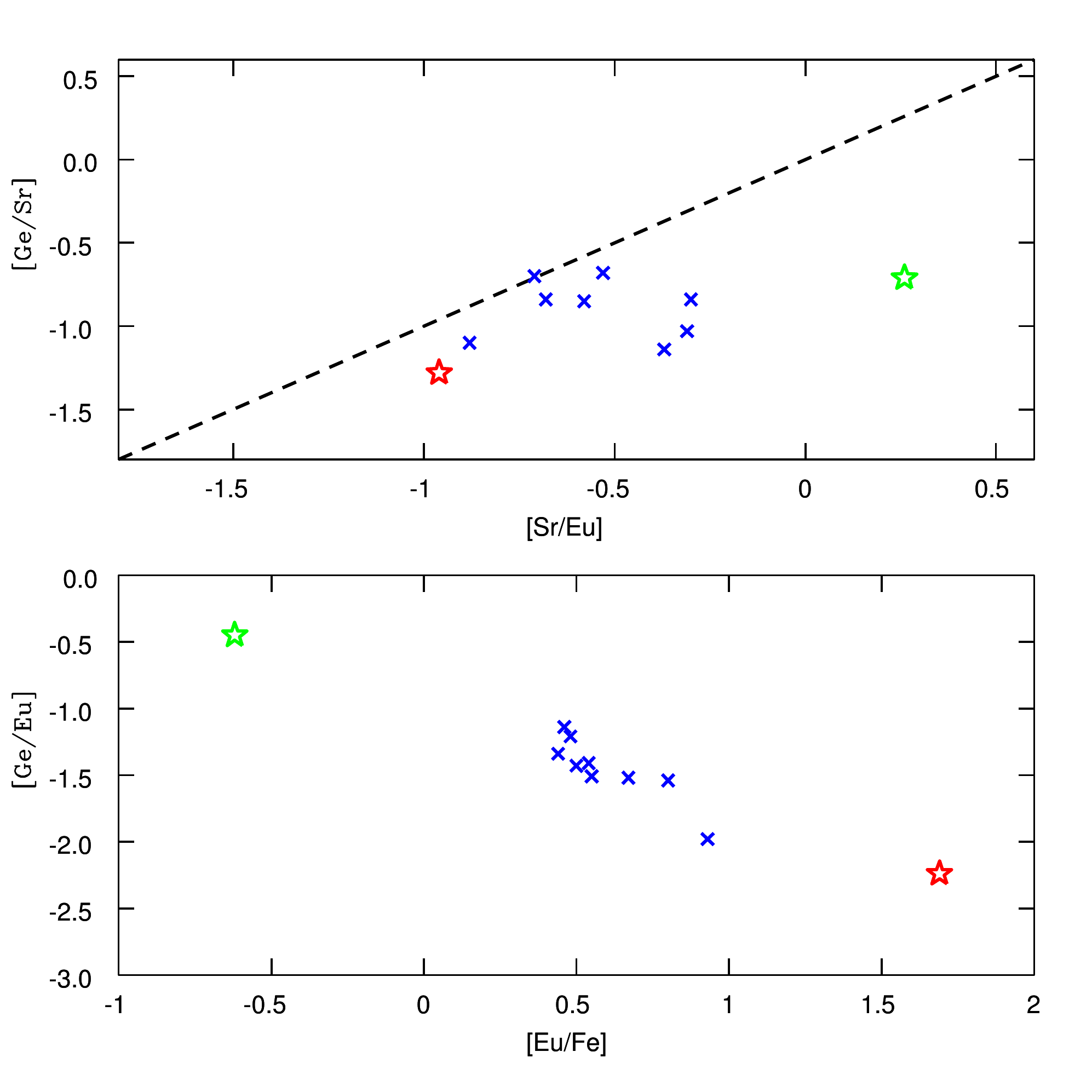}}
\caption{Relative abundances [Ge/Sr] displayed as a function of [Sr/Eu] (top) and correlation between the 
abundance ratios [Ge/Eu] and [Eu/Fe] (bottom). Symbols as in Fig. 19.}
\label{cowan2}
\end{figure}

It is important to note that the neutrino-driven wind always predicts Ge/Sr $<$ 1, because the high entropy (S $>$ 30 $k_{B}$/nucleon) leads to charged-particle freezeout from 
nuclear statistical equilibrium (NSE) and makes the abundance peak at N = 50 nuclei $^{88}$Sr, $^{89}$Y, and $^{90}$Zr 
\citep{Woosley1992,Meyer1998,WanajoIshimaru2006}. In contrast, the early convective ejecta from O-Ne-Mg (electron-capture) 
supernova predict Ge/Sr $\sim$ 1, because its low entropy (S $\sim$ 10 $k_{B}$/nucleon) with mild neutron-richness
 (Y$_{e}$ down to 0.4) leads to form the abundance 
peak at A = 70-80 (N $<$ 50) including Ge in NSE \citep{Hartmann1985,Wanajo2011}. In Fig. \ref{cowan2}
 we compare the relative abundances [Ge/Sr] with [Sr/Eu] for 
the same sample used in Fig. \ref{cowan1}, and one can find a marginal correlation between the Ge
 enhancement and the weakness of r-processing. 
The bottom panel in Fig. \ref{cowan2} still compares the 
abundance of Ge relative to the level of the heavy r-elements in the star, represented by
 the Eu abundance, as a function of the enhancement in r-process elements. 
We find a clear anticorrelation between the r-process richness and the Ge enhancement. 
In fact, our Fig. \ref{faro} shows that the low Ge abundance in our r-rich star CS 31082-001 is 
described better by the neutrino wind models, while the high Ge abundance in the r-poor star HD 122563 is explained
well by the electron-capture supernova model 
\citep[see Fig. 5 in][]{Wanajo2011}. This indicates that Ge serves as a key element in constraining the astrophysical
 conditions for r-process nucleosynthesis. 

\begin{figure}
%\resizebox{\hsize}{!}{\includegraphics[angle=0]{faro.pdf}}
\centering
\resizebox{85mm}{!}{\includegraphics{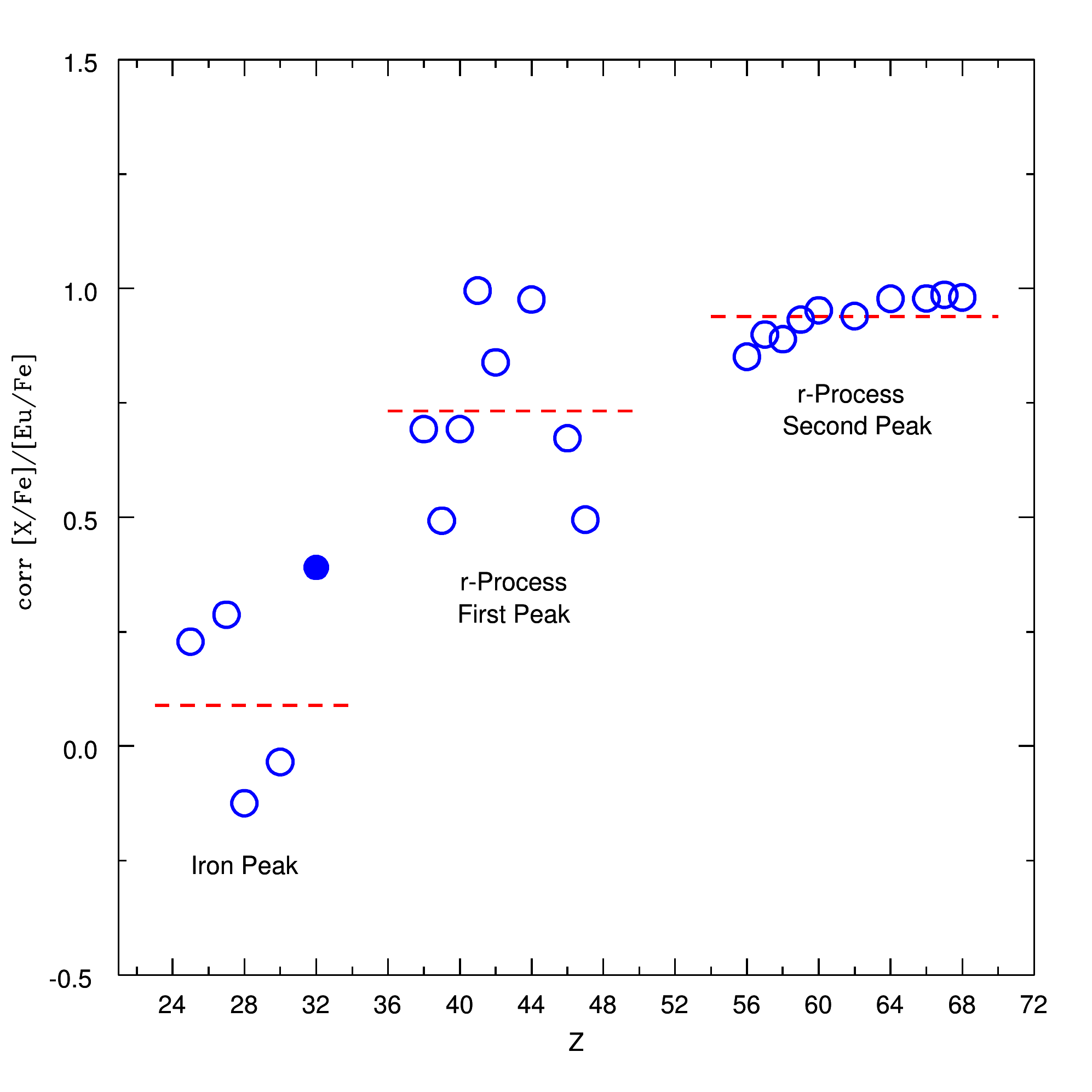}}
\caption{Correlation between [X/Fe] with respect to [Eu/Fe] using a sample of metal-poor stars (see text). The germanium is represented as the filled circle. 
The dashed lines represent the average behavior in each group of elements.}
\label{corr}
\end{figure}

It is worth noting that the region between the iron peak and the first peak of the r-process is historically 
thought to be the beginning of the r-process, and Ge is at the end of the Fe peak. 
In fact, the noncorrelation between the Ge abundance and the general level of heavy r-elements
 (Fig. \ref{cowan1}, bottom panel), as well as the anticorrelation 
between those (Fig. \ref{cowan2}, bottom panel), leads us not to discard the possibility of
 an iron peak (or NSE) origin to Ge. 
Using a sample of metal-poor stars from \citet{Cowan2005}, \citet{Francois2007}, \citet{Roederer2010b}, \citet{Peterson2011}, and \citet{Hansen2012}, we calculated the 
correlation between [X/Fe] with respect to [Eu/Fe], where X represents the elements 
available from the iron peak to the heavy r-elements. Figure \ref{corr} shows our 
results, and represents the average behavior in each group of elements, and in the case of the iron peak the value was calculated without the result for 
germanium. One can see that the behavior of the germanium is not clear enough to allow us to decide about its origin, and 
we stress that it is necessary to collect more observational data and perform an NLTE analysis of this element
 to permit a strong conclusion.

\section{Conclusions}

For the first time, we have determined abundances of molybdenum and germanium in CS 31082-001, using an STIS spectrum. 
In fact, there is a lack of results for the stellar abundances of these elements for EMP stars compared to other elements from the 
first peak, so our new abundance determinations are particularly valuable, and should be seen as part of a big project, 
from several authors, trying to solve this observational gap as an attempt to understand the origin of the lightest 
trans-Fe elements. 

Following the idea of completing the abundance analysis of this star, we were also able to 
derive abundances for lutetium 
and tantalum, other newly determined abundances, which agrees with the 
solar system r-process abundance, confirming the compatibility between the r-process in EMP 
stars and the solar system patterns, from Ba through the third r-process peak, which also 
confirms universal behavior of the process in this range of elements. 
On the other hand, this compatibility does not extend to the lighter heavy elements 
or to the heaviest elements from the actinides region.

Furthermore, the STIS/HST spectra of CS 31082-001 permitted us to derive the first abundance determination of W and Re in 
an EMP star, two extremely important elements for studying the transition region between the second and the third peaks of the 
r-process. Actually, together with the previous abundances, our new results make the star the most complete r-II object ever studied, 
and a major template for studies of r-process models, with a total of 37 detections of n-capture elements, supplanting the 
previous winner BD+17$^{\circ}$3248.

The abundances of the second peak of CS 31082-001 are reasonably well represented by those of the cold model by \citet{Wanajo2007}, but not the abundances of the first peak. 
As in some cases an additional production of the first peak elements by other processes could be invoked. More elegantly, the model of Farouqi with different electron 
abundances Y$_{e}$ can explain the first (including Ge) and second peaks. 
We also present the first NLTE+3D lead abundance in this star, which together with the other heavy elements from the third peak, 
lead us to underline that supernovae neutrino wind models with lower temperature satisfactorily describes the formation of the elements in this region. 
In general, the comparisons between calculations and observations do in fact argue for a combination of processes to reproduce the full range 
of observed stellar abundances.

\begin{acknowledgements}
CS and BB acknowledge grants from CAPES, CNPq and FAPESP. MS, VH, BP, RC, FS, PB and PF acknowledge the support of 
the CNRS (PNCG and PNPS). JA and BN acknowledge partial financial support from the Carlsberg Foundation and the Danish 
Natural Science Research Council. T.C.B. acknowledges partial funding of this work from grants PHY
02-16783 and PHY 08-22648: Physics Frontier Center/Joint Institute for Nuclear Astrophysics (JINA), awarded by the U.S. 
National Science Foundation.
\end{acknowledgements}

\begin{appendix}

\section{Line list and atomic data}

Table \ref{lines_total} presents the lines of neutron-capture elements that were used to derive abundances. 
The wavelengths, excitation potentials, and oscillator strengths are listed, together with references.

\onecolumn

\longtab{1}{
\begin{longtable}{cccccc}
\caption{\label{lines_total} Available spectral lines and abundances \textbf{obtained in CS 31082-001 using the spectra from HST/STIS and VLT/UVES}.}\\
\hline\hline
\hbox{$\lambda$({\rm \AA})} & \hbox{$\chi_{\rm ex}$ (eV)} & \hbox{log gf} & \hbox{A(X)$_{STIS}$} & \hbox{A(X)$_{UVES}$} & Ref. \\
\hline
\endfirsthead
\caption{continued.}\\
\hline\hline
\hbox{$\lambda$({\rm \AA})} & \hbox{$\chi_{\rm ex}$ (eV)} & \hbox{log gf} & \hbox{A(X)$_{STIS}$} & \hbox{A(X)$_{UVES}$} & Ref. \\
\hline
\endhead
\hline
\endfoot
\multicolumn{6}{c}{Ge I (Z=32)} \\
  3039.067 & 0.883 & -0.040 &  0.10 & -- & 1 \\
\multicolumn{6}{c}{Y II (Z=39)} \\
  3200.272 & 0.130 & -0.430 &  ---  & -0.08 & 2 \\
  3203.322 & 0.104 & -0.370 &  ---  & -0.07 & 2 \\
  3216.682 & 0.130 & -0.020 &  ---  & -0.27 & 2 \\
  3242.280 & 0.180 &  0.210 &  ---  & -0.10 & 2 \\
  3448.808 & 0.409 & -1.440 &  ---  & -0.15 & 2 \\
  3549.005 & 0.130 & -0.280 &  ---  & -0.12 & 2 \\
  3584.518 & 0.104 & -0.410 &  ---  & -0.13 & 2 \\
  3600.741 & 0.180 &  0.280 &  ---  & -0.20 & 2 \\
  3601.919 & 0.104 & -0.180 &  ---  & -0.18 & 2 \\
  3611.044 & 0.130 &  0.110 &  ---  & -0.23 & 2 \\
  3628.705 & 0.130 & -0.710 &  ---  & -0.13 & 2 \\
  3633.122 & 0.000 & -0.100 &  ---  & -0.25 & 2 \\
  3710.294 & 0.180 &  0.460 &  ---  & -0.11 & 2 \\
  3774.331 & 0.130 &  0.210 &  ---  & -0.14 & 2 \\
  3788.694 & 0.104 & -0.070 &  ---  & -0.13 & 2 \\
\multicolumn{6}{c}{Zr II (Z=40)} \\
  2699.593 & 0.039 & -1.170 &  0.73 & ---  & 3 \\
  2732.711 & 0.095 & -0.490 &  0.72 & ---  & 3 \\
  2758.792 & 0.000 & -0.560 & -0.07 & ---  & 3 \\
  2818.738 & 0.959 &  0.020 &  0.65 & ---  & 4 \\
  2915.973 & 0.466 & -0.500 &  0.69 & ---  & 3 \\
  2916.625 & 0.359 & -1.110 &  0.60 & ---  & 3 \\
  2952.236 & 0.164 & -1.250 &  0.65 & ---  & 4 \\
  2962.673 & 0.359 & -0.570 &  0.65 & ---  & 3 \\
  3019.832 & 0.039 & -1.130 &  0.80 & ---   & 4 \\
  3028.045 & 0.972 &  0.020 &  0.45 & ---   & 4 \\
  3030.915 & 0.000 & -1.040 &  0.66 &  0.60 & 4 \\ %CONFIRMAR DADOS
  3054.837 & 1.011 &  0.180 &  0.35 & ---   & 4 \\
  3061.334 & 0.095 & -1.380 &  0.35 & ---   & 4 \\
  3095.073 & 0.039 & -0.960 &  ---  &  0.50 & 4 \\
  3125.926 & 0.000 & -0.883 &  ---  &  0.69 & 4 \\
  3129.763 & 0.039 & -0.650 &  ---  &  0.63 & 4 \\
  3133.489 & 0.959 & -0.200 &  ---  &  0.58 & 4 \\
  3138.683 & 0.095 & -0.460 &  ---  &  0.57 & 4 \\
  3231.692 & 0.039 & -0.590 &  ---  &  0.70 & 4 \\
  3241.042 & 0.039 & -0.504 &  ---  &  0.39 & 4 \\
  3272.221 & 0.000 & -0.700 &  ---  &  0.70 & 4 \\
  3279.266 & 0.095 & -0.230 &  ---  &  0.60 & 4 \\
  3284.703 & 0.000 & -0.480 &  ---  &  0.65 & 4 \\
  3305.153 & 0.039 & -0.690 &  ---  &  0.68 & 4 \\
  3314.488 & 0.713 & -0.686 &  ---  &  0.50 & 4 \\
  3334.607 & 0.559 & -0.797 &  ---  &  0.65 & 4 \\
  3338.414 & 0.959 & -0.578 &  ---  &  0.40 & 4 \\
  3340.574 & 0.164 & -0.461 &  ---  &  0.43 & 4 \\
  3356.088 & 0.095 & -0.513 &  ---  &  0.59 & 4 \\
  3357.264 & 0.000 & -0.736 &  ---  &  0.70 & 4 \\
  3391.982 & 0.164 &  0.463 &  ---  &  0.70 & 4 \\
  3393.122 & 0.039 & -0.700 &  ---  &  0.60 & 4 \\
  3402.868 & 1.532 & -0.330 &  ---  &  0.55 & 4 \\
  3403.673 & 0.999 & -0.576 &  ---  &  0.45 & 4 \\
  3419.128 & 0.164 & -1.574 &  ---  &  0.50 & 4 \\
  3424.813 & 0.039 & -1.305 &  ---  &  0.62 & 4 \\
  3430.514 & 0.466 & -0.164 &  ---  &  0.54 & 4 \\
  3457.548 & 0.559 & -0.530 &  ---  &  0.68 & 4 \\
  3479.029 & 0.527 & -0.690 &  ---  &  0.58 & 5 \\
  3479.383 & 0.713 &  0.170 &  ---  &  0.20 & 5 \\
  3499.560 & 0.409 & -0.810 &  ---  &  0.44 & 5 \\
  3505.682 & 0.164 & -0.360 &  ---  &  0.60 & 5 \\
  3506.048 & 1.236 & -0.860 &  ---  &  0.43 & 4 \\
  3520.869 & 0.559 & -1.089 &  ---  &  0.20 & 4 \\
  3525.803 & 0.359 & -0.653 &  ---  &  0.25 & 4 \\
  3536.935 & 0.359 & -1.306 &  ---  &  0.35 & 4 \\
  3551.939 & 0.095 & -0.310 &  ---  &  0.65 & 5 \\
  3556.585 & 0.466 &  0.140 &  ---  &  0.00 & 4 \\
  3573.055 & 0.319 & -1.041 &  ---  &  0.63 & 4 \\
  3578.211 & 1.208 & -0.607 &  ---  &  0.52 & 4 \\
  3588.300 & 0.409 & -1.130 &  ---  &  0.60 & 5 \\
  3607.373 & 1.236 & -0.640 &  ---  &  0.35 & 5 \\
  3611.889 & 1.743 &  0.450 &  ---  &  0.25 & 4 \\
  3613.102 & 0.039 & -0.465 &  ---  &  0.67 & 4 \\
  3614.765 & 0.359 & -0.252 &  ---  &  0.54 & 4 \\
  3630.004 & 0.359 & -1.110 &  ---  &  0.52 & 5 \\
  3636.436 & 0.466 & -1.035 &  ---  &  0.52 & 4 \\
  3674.696 & 0.319 & -0.446 &  ---  &  0.30 & 4 \\
  3714.794 & 0.527 & -0.930 &  ---  &  0.72 & 5 \\
  3766.795 & 0.409 & -0.812 &  ---  &  0.67 & 4 \\
\multicolumn{6}{c}{Nb II (Z=41)} \\
  2876.957 & 0.439 & -0.490 & -0.62 & --- & 4 \\
  2908.237 & 0.292 & -0.340 & -0.62 & --- & 4 \\
  2910.581 & 0.376 & -0.190 & -0.70 & --- & 4 \\
  2911.738 & 0.326 & -0.270 & -0.62 & --- & 4 \\
  2950.878 & 0.514 &  0.210 & -0.50 & --- & 4 \\
  2994.718 & 0.514 & -0.250 & -0.15 & --- & 4 \\
  3028.433 & 0.439 & -0.410 & -0.38 & -0.27 & 4 \\
  3191.093 & 0.514 & -0.260 &  ---  & -0.55 & 4 \\
  3215.591 & 0.439 & -0.190 &  ---  & -0.58 & 4 \\
\multicolumn{6}{c}{Mo II (Z=42)} \\
  2660.576 & 1.492 & -0.136 & -0.15 & --- & 6 \\
  2871.507 & 1.540 &  0.056 & -0.26 & --- & 6 \\
  2930.485 & 1.492 & -0.228 &  0.08 & --- & 6 \\
\multicolumn{6}{c}{Ru I (Z=44)} \\
  2874.988 & 0.000 & -0.240 &  0.65 & ---   & 7 \\
  3436.736 & 0.148 &  0.015 &  ---  &  0.45 & 7 \\
  3498.942 & 0.000 &  0.310 &  ---  &  0.27 & 7 \\
  3728.025 & 0.000 &  0.270 &  ---  &  0.35 & 7 \\
\multicolumn{6}{c}{Rh I (Z=45)} \\
  3396.819 & 0.000 &  0.050 &  ---  & -0.45 & 4 \\
  3434.885 & 0.000 &  0.450 &  ---  & -0.41 & 4 \\
  3700.907 & 0.190 & -0.100 &  ---  & -0.40 & 4 \\
\multicolumn{6}{c}{Pd I (Z=46)} \\
  3242.700 & 0.814 & -0.070 &  ---  & -0.10 & 4 \\
  3404.579 & 0.814 &  0.320 &  ---  & -0.18 & 4 \\
  3516.944 & 0.962 & -0.240 &  ---  & -0.07 & 4 \\
  3634.690 & 0.814 &  0.090 &  ---  & -0.02 & 4 \\
\multicolumn{6}{c}{Ag I (Z=47)} \\
  3280.679 & 0.000 & -0.050 &  ---  & -1.03 & 4 \\
  3382.889 & 0.000 & -0.377 &  ---  & -0.65 & 4 \\
\multicolumn{6}{c}{Ce II (Z=58)} \\
  3263.885 & 0.459 & -0.390 &  ---  & -0.40 & 4 \\
  3426.205 & 0.122 & -0.660 &  ---  & -0.38 & 8 \\
  3507.941 & 0.175 & -0.960 &  ---  & -0.27 & 8 \\
  3520.520 & 0.175 & -0.910 &  ---  & -0.32 & 8 \\
  3534.045 & 0.521 & -0.140 &  ---  & -0.30 & 8 \\
  3539.079 & 0.320 & -0.270 &  ---  & -0.29 & 8 \\
  3577.456 & 0.470 &  0.140 &  ---  & -0.30 & 8 \\
  3659.225 & 0.175 & -0.670 &  ---  & -0.38 & 8 \\
  3709.929 & 0.122 & -0.260 &  ---  & -0.20 & 8 \\
  3781.616 & 0.529 & -0.260 &  ---  & -0.22 & 8 \\
\multicolumn{6}{c}{Nd II (Z=60)} \\
  3285.085 & 0.000 & -1.050 &  ---  & -0.08 & 4 \\
  3300.143 & 0.000 & -1.036 &  ---  & -0.30 & 4 \\
  3325.889 & 0.064 & -1.174 &  ---  & -0.20 & 4 \\
  3334.465 & 0.182 & -0.930 &  ---  & -0.22 & 9 \\
  3555.764 & 0.321 & -0.950 &  ---  & -0.30 & 9 \\
  3560.718 & 0.471 & -0.500 &  ---  & -0.38 & 9 \\
  3598.021 & 0.064 & -1.020 &  ---  & -0.22 & 9 \\
  3609.780 & 0.000 & -0.800 &  ---  & -0.25 & 9 \\
  3730.577 & 0.380 & -0.611 &  ---  & -0.15 & 4 \\
  3738.055 & 0.559 & -0.040 &  ---  & -0.21 & 9 \\
  3741.424 & 0.064 & -0.680 &  ---  & -0.15 & 9 \\
  3763.472 & 0.205 & -0.430 &  ---  & -0.20 & 9 \\
  3779.462 & 0.182 & -0.560 &  ---  & -0.26 & 9 \\
  3780.382 & 0.471 & -0.350 &  ---  & -0.28 & 9 \\
  3784.245 & 0.380 &  0.150 &  ---  & -0.13 & 9 \\
  3795.454 & 0.205 & -0.650 &  ---  & -0.21 & 9 \\
  3803.471 & 0.205 & -0.390 &  ---  & -0.20 & 9 \\
  3808.767 & 0.064 & -0.650 &  ---  & -0.12 & 9 \\
\multicolumn{6}{c}{Sm II (Z=62)} \\
  3218.596 & 0.185 & -0.640 &  ---  & -0.53 & 10 \\
  3244.686 & 0.185 & -1.330 &  ---  & -0.45 & 10 \\
  3253.403 & 0.104 & -0.770 &  ---  & -0.55 & 10 \\
  3304.517 & 0.000 & -1.190 &  ---  & -0.45 & 4 \\
  3307.027 & 0.659 & -0.301 &  ---  & -0.15 & 4 \\
  3321.189 & 0.378 & -0.362 &  ---  & -0.43 & 4 \\
  3384.654 & 0.378 & -0.741 &  ---  & -0.32 & 4 \\
  3568.271 & 0.485 &  0.290 &  ---  & -0.35 & 10 \\
  3583.372 & 0.185 & -1.110 &  ---  & -0.27 & 10 \\
  3604.281 & 0.485 & -0.030 &  ---  & -0.38 & 10 \\
  3609.492 & 0.277 &  0.160 &  ---  & -0.45 & 10 \\
  3621.210 & 0.104 & -0.510 &  ---  & -0.46 & 10 \\
  3627.004 & 0.277 & -0.510 &  ---  & -0.48 & 10 \\
  3661.352 & 0.041 & -0.360 &  ---  & -0.45 & 10 \\
  3670.821 & 0.104 & -0.240 &  ---  & -0.58 & 10 \\
  3706.752 & 0.485 & -0.600 &  ---  & -0.50 & 10 \\
  3718.883 & 0.378 & -0.310 &  ---  & -0.35 & 10 \\
  3731.263 & 0.104 & -0.330 &  ---  & -0.70 & 10 \\
  3739.120 & 0.041 & -0.430 &  ---  & -0.45 & 10 \\
  3743.877 & 0.333 & -0.550 &  ---  & -0.21 & 10 \\
  3758.460 & 0.000 & -1.102 &  ---  & -0.30 & 4 \\
  3760.710 & 0.185 & -0.400 &  ---  & -0.48 & 10 \\
  3762.588 & 0.248 & -0.850 &  ---  & -0.43 & 10 \\
\multicolumn{6}{c}{Eu II (Z=63)} \\
  2906.669 & 0.000 & -0.440 & -0.75 & --- & 11 \\
\multicolumn{6}{c}{Gd II (Z=64)} \\
  2833.748 & 0.492 & -0.096 & -0.22 & ---   & 4 \\
  3358.625 & 0.032 &  0.250 &  ---  & -0.32 & 12 \\
  3360.712 & 0.032 & -0.540 &  ---  & -0.33 & 12 \\
  3362.239 & 0.079 &  0.430 &  ---  & -0.30 & 12 \\
  3364.245 & 0.000 & -1.086 &  ---  & -0.35 & 4 \\
  3392.527 & 0.079 & -0.330 &  ---  & -0.25 & 12 \\
  3418.729 & 0.000 & -0.360 &  ---  & -0.22 & 12 \\
  3422.464 & 0.240 &  0.710 &  ---  & -0.06 & 12 \\
  3423.924 & 0.000 & -0.550 &  ---  & -0.34 & 12 \\
  3439.208 & 0.382 &  0.080 &  ---  & -0.36 & 12 \\
  3439.787 & 0.425 & -0.120 &  ---  & -0.28 & 12 \\
  3439.988 & 0.240 &  0.210 &  ---  & -0.24 & 12 \\
  3451.236 & 0.382 & -0.260 &  ---  & -0.32 & 12 \\
  3454.907 & 0.032 & -0.640 &  ---  & -0.29 & 12 \\
  3463.990 & 0.427 &  0.250 &  ---  & -0.32 & 12 \\
  3467.274 & 0.425 &  0.080 &  ---  & -0.39 & 12 \\
  3473.224 & 0.032 & -0.370 &  ---  & -0.23 & 12 \\
  3481.802 & 0.492 &  0.110 &  ---  & -0.35 & 12 \\
  3482.607 & 0.427 & -0.470 &  ---  & -0.35 & 12 \\
  3491.960 & 0.000 & -0.530 &  ---  & -0.25 & 12 \\
  3557.058 & 0.600 &  0.040 &  ---  & -0.28 & 12 \\
  3646.196 & 0.240 &  0.320 &  ---  & -0.39 & 12 \\
  3654.624 & 0.079 & -0.080 &  ---  & -0.27 & 12 \\
  3656.152 & 0.144 & -0.020 &  ---  & -0.36 & 12 \\
  3671.205 & 0.079 & -0.220 &  ---  & -0.25 & 12 \\
  3699.737 & 0.354 & -0.290 &  ---  & -0.37 & 12 \\
  3768.396 & 0.079 &  0.210 &  ---  & -0.25 & 12 \\
  3796.384 & 0.032 &  0.020 &  ---  & -0.22 & 12 \\
\multicolumn{6}{c}{Tb II (Z=65)} \\
  2934.802 & 0.126 & -0.596 & -0.50 & ---   & 4 \\
  3509.144 & 0.000 &  0.700 &  ---  & -1.05 & 13 \\
  3633.287 & 0.641 &  0.090 &  ---  & -1.00 & 13 \\
  3641.655 & 0.649 &  0.040 &  ---  & -1.00 & 13 \\
\multicolumn{6}{c}{Dy II (Z=66)} \\
  3407.796 & 0.000 &  0.180 &  ---  & -0.15 & 14 \\
  3413.784 & 0.103 & -0.520 &  ---  & -0.16 & 14 \\
  3434.369 & 0.000 & -0.450 &  ---  & -0.19 & 14 \\
  3454.317 & 0.103 & -0.140 &  ---  & -0.16 & 14 \\
  3456.559 & 0.590 & -0.110 &  ---  & -0.12 & 14 \\
  3460.969 & 0.000 & -0.070 &  ---  & -0.14 & 14 \\
  3531.707 & 0.000 &  0.770 &  ---  & +0.15 & 14 \\
  3534.960 & 0.103 & -0.040 &  ---  & -0.09 & 14 \\
  3536.019 & 0.538 &  0.530 &  ---  & -0.18 & 14 \\
  3546.832 & 0.103 & -0.550 &  ---  & -0.11 & 14 \\
  3550.218 & 0.590 &  0.270 &  ---  & -0.22 & 14 \\
  3563.148 & 0.103 & -0.360 &  ---  & -0.11 & 14 \\
  3694.810 & 0.103 & -0.110 &  ---  & -0.08 & 14 \\
\multicolumn{6}{c}{Er II (Z=68)} \\
  2897.518 & 1.654 &  0.573 & -0.20 & ---   & 4 \\
  2904.468 & 0.846 &  0.330 & -0.10 & ---   & 15 \\
  2964.520 & 0.846 &  0.580 & -0.30 & ---   & 15 \\
  3364.076 & 0.055 & -0.420 &  ---  & -0.40 & 15 \\
  3441.130 & 0.055 & -0.580 &  ---  & -0.30 & 15 \\
  3499.103 & 0.055 &  0.290 &  ---  & -0.40 & 15 \\
  3524.913 & 0.000 & -0.790 &  ---  & -0.40 & 15 \\
  3549.844 & 0.670 & -0.290 &  ---  & -0.36 & 15 \\
  3559.894 & 0.000 & -0.690 &  ---  & -0.45 & 15 \\
  3580.518 & 0.055 & -0.620 &  ---  & -0.35 & 15 \\
  3616.566 & 0.000 & -0.310 &  ---  & -0.14 & 15 \\
  3618.916 & 0.670 & -0.500 &  ---  & -0.12 & 15 \\
  3633.536 & 0.000 & -0.530 &  ---  & -0.46 & 15 \\
  3700.720 & 0.055 & -1.290 &  ---  & -0.22 & 4 \\
  3729.524 & 0.000 & -0.590 &  ---  & -0.29 & 15 \\
  3742.640 & 0.636 & -0.360 &  ---  & -0.36 & 15 \\
  3786.836 & 0.000 & -0.520 &  ---  & -0.34 & 15 \\
\multicolumn{6}{c}{Tm II (Z=69)} \\
  3015.294 & 0.029 & -0.590 & -1.00 & ---   & 16 \\
  3131.255 & 0.000 &  0.080 &  ---  & -1.25 & 16 \\
  3362.615 & 0.029 & -0.200 &  ---  & -1.00 & 16 \\
  3397.498 & 0.000 & -0.810 &  ---  & -1.11 & 16 \\
  3462.197 & 0.000 &  0.030 &  ---  & -1.31 & 14 \\
  3700.256 & 0.029 & -0.380 &  ---  & -1.18 & 14 \\
  3701.363 & 0.000 & -0.540 &  ---  & -1.29 & 14 \\
  3761.914 & 0.000 & -0.450 &  ---  & -1.22 & 16 \\
  3795.760 & 0.029 & -0.230 &  ---  & -1.22 & 14 \\
\multicolumn{6}{c}{Lu II (Z=71)} \\
  2847.505 & 1.463 & -0.230 & -1.03 & ---   & 17 \\
  2963.318 & 1.463 & -0.240 & -1.00 & ---   & 17 \\
  3077.605 & 1.542 &  0.160 &  ---  & -1.20 & 17 \\
\multicolumn{6}{c}{Hf II (Z=72)} \\
  3012.900 & 0.000 & -0.600 & -0.77 & ---   & 18 \\
  3109.113 & 0.787 & -0.260 &  ---  & -0.60 & 18 \\
  3255.279 & 0.452 & -1.210 &  ---  & -0.55 & 18 \\
  3399.793 & 0.000 & -0.570 &  ---  & -0.81 & 18 \\
  3569.034 & 0.787 & -0.460 &  ---  & -0.90 & 18 \\
\multicolumn{6}{c}{Ta II (Z=73)} \\
  2635.583 & 0.128 &  0.700 & -2.15 & --- & 4 \\
  2832.702 & 0.847 & -0.070 & -1.05 & --- & 4 \\
\multicolumn{6}{c}{W II (Z=74)} \\
  2697.706 & 0.188 & -0.870 & -0.90 & --- & 4 \\
\multicolumn{6}{c}{Re I (Z=75)} \\
  2930.613 & 1.867 &  2.000 & -0.20 & --- & 4 \\
\multicolumn{6}{c}{Re II (Z=75)} \\
  2637.006 & 2.373 &  1.020 & -0.21 & --- & 4 \\
\end{longtable}
\tablebib{(1)~\citet{Biemont1999}; (2)~\citet{Hannaford1982}; (3)~\citet{Ljung2006}; (4)~VALD; (5)~\citet{Biemont1981}; 
(6)~\citet{Sikstrom2001}; (7)~\citet{Wickliffe1994}; (8)~\citet{Lawler2009}; (9)~\citet{Den Hartog2003}; (10)~\citet{Lawler2006}; 
(11)~\citet{Zhiguo2000}; (12)~\citet{Den Hartog2006}; (13)~\citet{Lawler2001c}; (14)~\citet{Sneden2009}; (15)~\citet{Lawler2008}; 
(16)~\citet{Wickliffe1997}; (17)~\citet{Quinet1999}; (18)~\citet{Lawler2007}}
}

\end{appendix}
\end{document}